\documentstyle[aps, amsfonts, multicol,preprint]{revtex}

\def\gtrless{\raise2.5pt\hbox{$>$}\llap{\lower2.5pt\hbox{$<$}}}

\begin{document} 

\draft

\title{Reorientational relaxation of a\\ linear 
probe molecule in a simple glassy liquid} 
\author{W.~G{\"o}tze, A. P. Singh, and Th. Voigtmann} 
\address{Physik-Department, Technische Universit{\"a}t M{\"u}nchen, 
85747 Garching, Germany} 
\date{Received 29 December 1999} 
 
\maketitle 
\begin{abstract} 
Within the mode-coupling theory (MCT) for the evolution  
of structural relaxation in glass-forming liquids, 
correlation functions and susceptibility spectra 
are calculated characterizing 
the rotational dynamics of a top-down symmetric dumbbell 
molecule, consisting of two fused hard spheres immersed in a 
hard-sphere system. It is found that for sufficiently 
large dumbbell elongations, the dynamics of the probe molecule follows 
the same universal glass-transition scenario as known from the  
MCT results of simple liquids. The $\alpha$-relaxation process of 
the angular-index-$j$=1 response is stronger, slower and less 
stretched than the one for $j$=2, in qualitative agreement with results 
found by dielectric-loss and depolarized-light-scattering 
spectroscopy for some supercooled liquids.  
For sufficiently small elongations, the 
reorientational relaxation occurs via large-angle flips, and the 
standard scenario for the glass-transition dynamics is 
modified for odd-$j$ responses due to precursor phenomena of a 
nearby type-A MCT transition. 
In this case, a major part of the relaxation outside the transient 
regime is described qualitatively by the $\beta$-relaxation scaling 
laws, while the $\alpha$-relaxation scaling law is strongly disturbed. 
\end{abstract} 
 
\section{Introduction} 
During the past ten years, the evolution of structural relaxation 
in glass-forming liquids has been intensively studied using 
neutron-scattering spectroscopy, various light-scattering  
techniques, dielectric-loss spectroscopy, and 
molecular-dynamics simulation.
Results of this work have also been used to test the 
mode-coupling theory (MCT), which interpretes the structural 
relaxation as precursor of the glass-transition.
Originally, the MCT was proposed as an approximation approach 
for the cage effect in liquids \cite{Leutheusser84,Bengtzelius84}. 
In its simplest version, the MCT equations of motion describe an 
ideal liquid-to-glass transition, i.e. a bifurcation 
from ergodic to nonergodic dynamics, if 
control parameters like temperature $T$ or packing fraction 
$\varphi$ cross critical values 
$T_c$ or $\varphi_c$, respectively. 
This bifurcation is connected with the evolution of a two-step 
relaxation scenario entirely determined by the regularly changing 
equilibrium structure. 
Two divergent time scales appear, closely connected to two 
power-law decay processes. A detailed description of these results can be
found in Ref. \cite{Goetze92} and references therein. 
Comparisons of the theoretical results for simple model systems with 
experiments done on colloids \cite{Megen93b,Megen95}, and with 
computer-simulation studies \cite{Gleim98,Gleim99}  
demonstrate the validity of the microscopic MCT approach. 
For the solutions of the MCT equations, 
a variety of results has been derived 
by asymptotic expansions, using as a small parameter the distance 
from the critical point, 
$\epsilon=(\varphi-\varphi_c)/\varphi_c$, or $\epsilon=(T_c-T)/T_c$, 
respectively. The leading-order results of this expansion 
establish universality features of the MCT dynamics. 
Assessments of the theory have been 
reached by comparing spectra in the GHz regime 
or relaxation curves within the pico-second window with the 
universal results. The outcome of this work, 
which is reviewed in Ref. \cite{Goetze99}, leads to the 
conclusion that MCT properly describes some essential  
features of structural relaxation even for some complicated 
molecular liquids. 
 
The MCT for simple systems has been extended recently 
to liquids of nonspherical molecules 
\cite{Schilling97,Franosch97c,Fabbian99b}. 
But so far, only 
the bifurcation equation for the so-called 
nonergodicity parameters resulting within the new theory  
could be solved. 
Comparing these results 
with the findings of molecular-dynamics simulations for a 
liquid of linear molecules \cite{Theis98,Fabbian98b} and 
for water \cite{Fabbian99b} 
indicates that the MCT for molecular liquids 
is promising. It was also predicted that there can be two 
states of nonergodic motion for nonspherical 
molecules. These states are connected by 
a type-A transition if the molecules exhibit a 
top-down symmetry 
\cite{Schilling97,Franosch97c,Franosch98b}. Such transitions are 
generic possibilities in MCT, provided there is some symmetry 
in the problem rendering certain mode-coupling coefficients 
zero \cite{Franosch94}. At a 
type-A transition, the nonergodicity parameters change 
continuously, whereas at the conventional MCT transition, 
referred to in this context as a type-B transition, 
a discontinuity occurs \cite{Goetze91b}. 
 
In this paper, correlation functions and susceptibility spectra 
shall be discussed, which 
deal with the glassy dynamics of the orientational degrees of 
freedom of nonspherical molecules. The results are obtained 
as solutions of the equations of 
motion derived previously \cite{Franosch97c} for the dynamics of 
a linear probe molecule immersed in a simple liquid. A top-down 
symmetric dumbbell of two fused hard spheres will be 
considered as the molecule, and as the solute, a hard-sphere 
system is chosen. This model deals with the simplest problem concerning 
glassy rotational dynamics, namely the influence of 
the cages formed by the neighbors of the molecule on 
the molecule's reorientational motion as it is caused by steric 
hindrance. The dynamics will be exemplified for two cases: a 
molecule with a large elongation and a molecule with a small 
elongation.  
 
It will be shown that large elongations lead to strong 
coupling of the rotational degrees of freedom to the density 
fluctuations of the solute, such that the glassy dynamics of 
the latter enforces the validity of all the universal MCT laws 
for the solvent. Moreover, the 
corrections to the leading-order-asymptotic laws show the same 
qualitative trends as studied for simple liquids 
\cite{Franosch97,Fuchs98}. A motivation of the present 
study is the explanation of three general 
properties of the $\alpha$-relaxation in molecular liquids, 
which are exhibited in Fig. \ref{ersteAbb}. In this figure, 
experimental susceptibility spectra for the van-der-Waals liquid 
propylene carbonate (PC) are reproduced for four 
temperatures. One set of data deals with the response for 
angular-momentum index $j$=1; it was obtained by 
dielectric-loss spectroscopy \cite{Schneider99}. The other 
set was measured by depolarized-light-scattering 
spectroscopy \cite{Du94} and deals with the ($j$=2) 
reorientational dynamics. The data show for $T=293$K and $T=295$K 
$\alpha$-relaxation peaks at $4$GHz ($j$=1) and $10$GHz ($j$=2) 
respectively. These temperatures exceed the 
melting temperature $T_m=218$K of PC by more than 
$70$K. Lowering $T$ to $200$K, the $\alpha$-peaks  
of the spectra are shifted down by about two orders of magnitude. 
The shape of the $\alpha$-peak 
is temperature independent, and the ratio of the 
$\alpha$-process-time scales, characterizing the 
$\alpha$-peak-maximum positions for the two values of $j$, 
is also $T$-independent. 
These are two features which MCT predicts to be universal.
The first nonuniversal feature to be understood is, that the  
$\alpha$-peak intensity, taken relative to that of the band of
microscopic excitations at around 1THz, is larger for the 
($j$=1) response than for the ($j$=2) case: 
the former exceeds the latter by about a factor of $2.7$. 
Second, the ($j$=1) response is slower than the response for $j$=2: 
the ratio of the $\alpha$-peak positions is about $2.5$. Third,  
the $\alpha$-peak of the ($j$=1) response is less stretched than the  
peak for $j$=2, i.e. the halfwidth of the ($j$=1) peak is smaller than 
that of the ($j$=2) peak. If one describes these peaks by the 
spectra of the Kohlrausch law, 
$\Phi(t)\propto e^{-(t/\tau)^{\beta}}$, the stretching 
exponent $\beta$ for $j$=1, 
$\beta_{j=1}\approx0.9$ \cite{Schneider99}, 
is larger than the  
one for $j$=2, $\beta_{j=2}\approx0.8$ \cite{Du94}.  
The same three $\alpha$-peak features are noticed, if one compares 
the depolarized-light-scattering spectra of glycerol \cite{Wuttke94} 
with the corresponding dielectric-loss spectra \cite{Lunkenheimer96}.
A fourth general feature to be explained is the large ratio of the 
$\alpha$-relaxation-time scale found by depolarized-light-scattering 
spectroscopy and the one found for the longitudinal elastic modulus 
by Brillouin-scattering spectroscopy.
For Salol, a ratio of about 10 was reported \cite{Dreyfus96}, while for 
PC, a factor of about 5 was found \cite{Du94}.

The small elongation of concern in this paper 
is chosen so that it exceeds the critical value 
for the above-mentioned type-A transition by about $10$\%. The theory 
for the corrections to the leading-order asymptotic laws
\cite{Franosch97,Fuchs98}
implies that these 
diverge at a type-A transition. Therefore, the range 
of validity of the universality features of the standard MCT bifurcation 
shrinks upon approaching the type-A transition. It will be shown  
that in our example the standard results are not exhibited any more for 
reasonable choices of the distance parameter $\epsilon$. 
In particular, it is impossible to identify a two-step scenario for  
the odd-$l$ correlators, nor is $\alpha$-relaxation scaling  
observed. 
 
The paper is organized as follows. In Sec. \ref{Themodelsystem}, the 
model is defined, and the  
MCT equations are noted. 
After an overview of the general scenario 
for the evolution of the glassy relaxation of the reorientational 
correlators (Sec. IIIA), the differences between the relaxation 
patterns for the ($j$=1) and ($j$=2) response are described for 
strong (Sec. IIIB) and weak (Sec. IIIC) steric hindrance.
In Sec. IIID it is demonstrated how the $\beta$-relaxation is described 
by the first scaling law, and in Sec. IIIE it is discussed how the 
$\alpha$-relaxation-scaling-law description emerges. 
The concluding Sec. IV summarizes the results.

\section{The Model system} 
\label{Themodelsystem} 
\subsection{The solvent} 
\label{eqnsolv} 
 
A system of $N$ spherical particles shall be considered as the solvent. 
The basic variables describing the structure are the density  
fluctuations for the wave vector $\vec{q}$:  
$\varrho_{\vec{q}} 
= \sum_\kappa \exp({\rm i}\vec{q}\cdot \vec{r}^{\,\kappa})/\sqrt{N}$. 
Here $\vec{r}^{\,\kappa}, \kappa =1,2,.., N$, labels the centers of the 
particles. The structure factor   
$S_q = \langle | \varrho_{\vec{q}}|^2 \rangle$ provides the simplest  
information on the equilibrium distribution of the particles; here,  
$\langle \,  \rangle$ denotes canonical averaging. Because of 
rotational symmetry, $S_q$ depends on the wave-vector 
modulus $q=|\vec{q}|$ only. The structure factor can be expressed 
through the direct correlation function $c_q$ via the Ornstein-Zernicke 
equation $S_q=1/(1-\rho c_q)$; where $\rho$ denotes the particle  
density \cite{Hansen86}. 
The simplest quantities, characterizing the structural dynamics in a
statistical manner, are the normalized auto-correlation functions for
the density fluctuations, called the density correlators 
$\Phi_q(t) = \langle \varrho_{\vec{q}}(t)^* \varrho_{\vec{q}}\rangle 
/ S_q$. The evolution with increasing time $t$ is given by the canonical
equations of motion. We will also need 
Fourier-Laplace transforms for complex frequency $z$, 
${\rm Im} z \geq 0$, $\Phi_q(z)$, using the convention: 
$F(z)={\rm i} \int_0^{\infty} \exp({\rm i} z t)~ F(t) {\rm d}t$. For
real frequency $\omega$, one gets with $z=\omega+{\rm i}0$:
$F(z)=F^{\prime}(\omega)+{\rm i}F^{\prime\prime}(\omega)$. The imaginary 
part $F^{\prime\prime}(\omega)$ is called the fluctuation spectrum, and
$\chi^{\prime\prime}(\omega)=\omega F^{\prime\prime}(\omega)$ is the
susceptibility spectrum \cite{Hansen86}.

The basic version of MCT consists of two equations
\cite{Bengtzelius84}. The first one is exact and derived within
the Zwanzig-Mori formalism:
\begin{mathletters}
\label{eins}
\begin{equation} 
\partial _{t}^{2}{\Phi }_{q}(t)+\Omega _{q}^{2}\Phi _{q}(t)+ 
\Omega _{q}^{2} \int_{0}^{t}~ m_{q}(t-t^{\prime })~ 
\partial _{t^{\prime }}{\Phi }_{q}(t^{\prime })~{\rm d}t^{\prime 
}=0\,.  \label{einsa} 
\end{equation} 
Here, $\Omega _{q}=vq/\sqrt{S_q}$, with $v$ denoting the thermal
velocity. The
relaxation kernel $m_q(t)$ is a fluctuating-force correlator. The
equation has to be solved with the initial condition 
$\Phi_{q}(t)=1-(\Omega_qt)^2/2+{\cal O}(t^3)$ \cite{Hansen86}. 
Equation (\ref{einsa}) is equivalent to the double fraction
\begin{equation}
\Phi_q(z)=\frac{\displaystyle -1}{\displaystyle z-
\frac{\Omega _{q}^{2}}{z+\Omega
_{q}^{2}~ m_q(z)}}.
\label{einsb}
\end{equation}
\end{mathletters}
The second MCT equation is obtained by writing the kernel as a sum of a
regular term and a
contribution describing the cage effect. The latter is treated by
Kawasaki's factorization approximation for the force correlations. It is
found to be a quadratic functional of the density fluctuations:
$\sum_{\vec{k}+\vec{p}=\vec{q}}~ V(\vec{q},\vec{k},\vec{p})\, 
\Phi_{k}(t)\,\Phi_{p}(t)$. 
For the sake of simplicity, the regular term shall be neglected in the
following. Furthermore, the wave-vector modulus will be discretized to
$M$ values with equal spacing $h$. Thus, $q$, $k$, $p$ can be considered
as labels running from $1$ to $M$. As a result, the kernel is 
given as a quadratic mode-coupling polynomial ${\cal F}_q$ 
of the $M$ correlators $\Phi_q(t)$, $q=1, ...,M$:
\begin{equation} 
m_{q}(t)={\cal {F}}_{q}[\Phi_k (t)]=\sum_{kp}
V_{qkp}\,\Phi_{k}(t)\,\Phi_{p}(t).
\label{zwei} 
\end{equation} 
The positive coupling coefficients $V_{qkp}$ are given by $S_q$ 
and $c_q$ \cite{Franosch97}. Anticipating these
equilibrium quantities to be known, Eqs. (\ref{einsa}) and
(\ref{zwei}) are closed.

Equations (\ref{eins}) and (\ref{zwei}) exhibit a transition from
liquid-state dynamics in the regime $T>T_c$ or $\varphi<\varphi_c$
to glass-state dynamics for $T\leq T_c$ or $\varphi\geq\varphi_c$. 
In the former regime the density 
fluctuations decay to zero
for long times, $\Phi_q(t \to \infty)=0$. The ideal-glass states
exhibit a nontrivial long-time limit, which is called the
nonergodicity parameter, $f_q=\Phi_q(t\to\infty)>0$. It is the
Debye-Waller factor of the glass. At the transition, 
this long-time limit
is discontinuous, and the jump is called critical nonergodicity parameter
or plateau, $f_q^c=f_q(T\nearrow T_{c}, \varphi\searrow\varphi_{c})>0$.
At the critical point, the correlators decay algebraically:
$\Phi_q(t)=f_q^c+h_q(t/t_0)^{-a}+{\cal O}((t/t_0)^{-2a})$. The 
exponent $a$, $0<a<1/2$, is called the critical exponent, and $h_q>0$ is
denoted as the critical amplitude. $t_0$ marks the time scale of the 
transient from the microscopic motion to the relaxation dynamics of the MCT. 
The MCT $\alpha$-process is defined
as the dynamics for those times, where
the correlators of the liquid decay from the plateau $f_q^c$ to
zero. The MCT $\beta$-process deals with the dynamics, where the
correlators are near the plateau, i.e. $|\Phi_q(t)-f_q^c|\ll 1$.  The first
relaxation step of the anomalous dynamics is given by the initial part
of the $\beta$-process; it deals with the decay towards the plateau for
times outside the transient: $(t/t_0)\gg1$, $\Phi_q(t)\geq f_q^c$. The
second step is the $\alpha$-decay in the liquid.  Its initial part
is identical with the final part of the $\beta$-process, and it follows
von Schweidler's law $\Phi_q(t)-f_q^c\propto -h_q t^b$.
The exponent $b$, $0<b\leq 1$, is called the von-Schweidler
exponent.

In a leading-order expansion in the small parameter $|\Phi_q(t)-f_q^c|$ 
one finds the universal results for the $\beta$-process. There holds
the factorization theorem 
\begin{equation} 
\label{drei} 
\Phi_q(t)-f_q^c=h_q~ G(t). 
\end{equation} 
The dependence on time and on control parameters is given by the
$q$-independent function $G(t)$, which is called the
$\beta$-correlator. It is determined by the equation
\begin{equation} 
\label{viera} 
\sigma-\lambda G(t)^2=\frac{{\rm d}}{{\rm d} t} \int_0^t~ G(t-t^{\prime}) 
G(t^{\prime})~ {\rm d}t^{\prime} ,
\end{equation} 
to be solved with the initial condition $G(t \to 0)=(t/t_0)^{-a}+{\cal
O}(t^a)$. The number $\lambda$, $0<\lambda<1$, is referred to as 
the exponent parameter. 
$\sigma$ is a smooth function of the control parameters and is
called the separation parameter. Its zero defines the critical
point. Expanding in leading order in the distance $\epsilon$, 
one can write $\sigma=C\epsilon\, , \, C>0$.

From Eq. (\ref{viera}), one derives the first scaling law
\begin{equation}
\label{fuenf}
G(t)=c_{\sigma}~ g_{\pm}(\hat{t})\, , \epsilon \gtrless 0 \qquad 
\hat{t}=t/t_{\sigma}.
\end{equation}
Here, $c_{\sigma}=\sqrt{|\sigma|}$ denotes the amplitude scale, and
$t_{\sigma}$ abbreviates the first characteristic time scale of the
MCT-transition scenario:
\begin{mathletters}
\label{sechs}
\begin{equation}
\label{sechsa}
t_{\sigma}=t_0/|\sigma|^{1/2a}.
\end{equation}
The master functions 
$g_{\pm}(\hat{t})$ are determined by $\lambda$ as solutions of
Eq. (\ref{viera}) for $\sigma=\pm 1$, respectively. They interpolate
monotonously between $g_{\pm}(\hat{t}\ll 1)=\hat{t}^{-a}$ and 
$g_{+}(\hat{t}\gg 1)=1/\sqrt{1-\lambda}$ or $g_{-}(\hat{t}\gg
1)=-B\hat{t}^{b}$. Von Schweidler's law is obtained as the
long-time limit on the scale $t_{\sigma}$ in the form: 
$\Phi_q(t)=f_q^c-h_q(t/t_{\sigma}^{\prime})^b$. Here $t_{\sigma}^{\prime}$
abbreviates the second characteristic scale of the theory:
\begin{equation} 
\label{sechsb}
t_{\sigma}^{\prime}=B^{-1/b} t_0 /|\sigma|^{\gamma}\, , \qquad 
\gamma=(1/2a)+(1/2b).
\end{equation}
\end{mathletters}
The $\alpha$-process obeys for $\epsilon\to 0$ the second scaling-law, 
called the superposition principle:
\begin{equation}
\label{sieben}
\Phi_q(t)=\tilde{\Phi}_q(\tilde{t})\, , 
\qquad \tilde{t}=t/t_{\sigma}^{\prime}.
\end{equation}
The control-parameter independent master function 
$\tilde{\Phi}_q(\tilde{t})$ exhibits the initial
decay $\tilde{\Phi}_q(\tilde{t})=f_q^c-h_q\tilde{t}^b+{\cal
O}(\tilde{t}^{2b})$. The parameters $f_q^c$, $h_q$ and $\lambda$ are
determined by ${\cal F}_q$ from
Eq. (\ref{zwei}) for control parameters at the critical point. The same
holds for the function $\tilde{\Phi}_q(\tilde{t})$. The constant $C$ is
determined by the first Taylor coefficient in $\epsilon$ of 
the deviations of ${\cal F}_q$ from its value at the critical
point. Formulas for these quantities can be found in
Ref. \cite{Goetze91b}, where also the original work is cited. 
The theory for the leading corrections to the quoted results has
been worked out in Ref. \cite{Franosch97}. 

The calculations in this paper will be done for the hard-sphere system
(HSS). The temperature does not enter the structure, but determines the
time scale via the thermal velocity only: $v^2\propto T$. 
The relevant control parameter is the packing fraction: 
$\varphi=\pi (\rho d^3)/6$, 
where $d$ is the particle diameter. The structure factor
will be calculated within the Percus-Yevick theory \cite{Hansen86}. The
discretization will be done for $M=100$ wave-vector values with step
size $hd=0.4$. For this model, all the mentioned MCT quantities have been
reported in Ref. \cite{Franosch97}. In particular it was found:
$\varphi_c=0.516$, $C=1.54$,
\begin{equation}
\label{neun}
\lambda=0.735, \qquad a=0.312, \qquad b=0.583, 
\qquad \gamma=2.46, \qquad B=0.836 .
\end{equation}
The results for the glass-transition of the HSS are documented 
comprehensively in Refs. \cite{Franosch97,Fuchs98}, albeit for a Brownian
microscopic dynamics. The bifurcation scenario for the model with
Newtonian dynamics as defined by Eqs. (\ref{eins}) and 
(\ref{zwei}) is demonstrated for the wave-vector $q=10.6/d$ in
Ref. \cite{Franosch98}, where the transient time-scale was
determined: $t_{0}=0.0236 (d/v)$.
For the presentation of our results in the following figures, the units
of length and time will be chosen so that $d=1$ and $d/v=1$. The
control parameters $\varphi$ shall be cited by the logarithm $x$ of the
distance parameter $\epsilon$:
\begin{equation}
\label{elf}
(\varphi-\varphi_c)/\varphi_c=\epsilon=\pm 10^{-x}\,.
\end{equation} 
As in the previous work \cite{Franosch97,Fuchs98,Franosch98}, 
the MCT equations are solved in the time domain. The solutions are 
then Laplace-transformed to get 
$\Phi_q^{\prime}(\omega)+{\rm i}\Phi_q^{\prime\prime}(\omega)$. 
Similarly, the transformed kernel $m_q^{\prime}(\omega)+{\rm i} 
m_q^{\prime\prime}(\omega)$ is calculated from $m_q(t)$ in Eq. (\ref{zwei}). 
These results are used to compare the left-hand side of 
Eq. (\ref{einsb}) with the right-hand side. 
Thereby a verification of the numerical solutions is obtained. 
 
\subsection{The solute} 
\label{eqnsolu} 
As a model for a dilute solution of molecules we shall consider a single 
linear molecule immersed in a simple system. The position of this
molecule is described by the tensor-density fluctuations 
$\varrho_{j}^{\nu}(\vec{q})={\mathrm{R}}_{j}^{\nu}(\vec{e})
\exp({\mathrm{i}}\vec{q}\cdot \vec{r})$. 
Here, $\vec{r}$ denotes the center-of-mass position and
$\vec{e}$ abbreviates the axis of the molecule. The
${\mathrm{R}}_{j}^{\nu}$ are related to the spherical harmonics by: 
${\mathrm{R}}_{j}^{\nu}(\vec{e})={\rm i}^j~ \sqrt{4\pi}~ {\rm
Y}_j^{\nu}(\vec{e})$. The solute-solvent equilibrium correlations are
described by the generalized structure factors 
$S_{J}(q)=\left\langle \varrho^{\ast}(\vec{q}_{0})\varrho_{J}^{0}(\vec 
{q}_{0})\right\rangle$, where $\vec{q}_0=(0,0,q)$. The proper
generalization of the density correlators for simple systems are
tensor-density correlators for the molecule,  
$\langle\varrho_{i}^{\mu}(\vec{q}_0,t)^{\ast}
\varrho_{j}^{\mu}(\vec{q}_0)\rangle $. The MCT for these quantities shall
be simplified by restricting the correlators to the diagonal elements
\begin{equation}
\label{zwoelf}
\Phi(q j \mu,t)=\langle\varrho_{j}^{\mu}(\vec{q}_0,t)^{\ast}
\varrho_{j}^{\mu}(\vec{q}_0)\rangle .
\end{equation}
Correlation functions for wave vectors $\vec{q}$ different from $\vec{q_0}$ 
can be obtained from the specified ones by elementary transformations 
\cite{Franosch97c}.

The first equation of the MCT for the molecule dynamics reads
\cite{Franosch97c}:
\begin{equation} 
\Phi (qj\mu ,z)=\frac{-1}{\displaystyle z-
\frac{\Omega_{{\rm {T}}q}^{2}}{z+\Omega _{{\rm {T}} q}^{2}(q)~ 
m_{{\rm T}}(qj\mu ,z)}-\frac{\Omega_{{\rm {R}} 
 j}^{2}}{z+\Omega _{{\rm {R}} j}^{2}~ m_{{\rm R}}(qj\mu ,z)}} .
\label{dreizehn} 
\end{equation} 
Here, $\Omega_{{\rm {T}},q}=vq$ is the
characteristic frequency for the translational motion of a tagged
particle. $\Omega_{{\rm {R}},j}=v_{\rm R}\sqrt{j(j+1)}$ is the
analogue for the rotational dynamics, where $j(j+1)$ plays 
here and in the following 
a similar role as $q^2$ for the translational motion.
The frequency $v_{\rm R}$ denotes the
thermal velocity for the rotation. The relaxation 
kernels $m_{{\rm T}}$ and $m_{{\rm R}}$ are
approximated along the same lines as indicated above for simple systems. 
They are obtained as a functional of the density
correlators of the solvent, multiplied by the tensor-density correlators
of the solute \cite{Franosch97c}. 
Let us discretize the wave vector to, say, $M^{\prime}$
values with equal spacing $h^{\prime}$. Let us also restrict the
angular-momentum index by some upper cutoff value $l_{co}$. One obtains 
the kernels as mode-coupling polynomials
\begin{eqnarray}
\nonumber
m_{\alpha }(qj\mu ,t)&=&{\cal F}_{\alpha qj\mu }\left[ \Phi (kl\nu ,t),
\Phi_{p}(t)\right]\\
\label{vierzehn}
&=&\sum_{kpl\nu } V_{\alpha qj\mu}(kpl\nu )~ 
\Phi(kl\nu, t)~ \Phi_{p}(t)\, , \qquad \alpha={\rm R,T} .
\end{eqnarray}
The positive coefficients $V_{\alpha qj\mu}(kpl\nu)$, 
$j,l=0,1,..,l_{co}$ are given in Ref. \cite{Singh99} 
as specialization of the results in Ref. \cite{Franosch97c}.
They are expressed in terms of $S_q$ and $S_{J}(q)$ for
$J=0,1,...,2l_{co}$. 
Anticipating $S_q$, $S_J(q)$ and $\Phi_q(t)$ as known, Eq. 
(\ref{dreizehn}) and (\ref{vierzehn}) are closed 
equations for the determination of the
$M^{\prime}\cdot (l_{co}+1)^{2}$ correlators $\Phi(qj\mu,t)$. 

The quantities of main interest for a statistical description of the
rotation of the molecule are the reorientational correlators, 
defined with the Legendre polynomials ${\rm P}_j$: 
\begin{mathletters}
\label{fuenfzehnneu}
\begin{equation} 
C^{(j)}(t)=\left\langle {\mathrm{P}}_{j}
\left( \vec{e}(t)\cdot \vec{e}\right) 
\right\rangle , \qquad j=1,2, ...\qquad .
\label{fuenfzehn} 
\end{equation} 
They are the long-wave-length limits of the general
correlators; $\Phi(q\to 0~ j0,t)=C^{(j)}(t)$ \cite{Franosch97c}. 
One gets from 
Eq. (\ref{dreizehn}) the fraction representation in analogy to
Eq. (\ref{einsb}):
\begin{equation} 
C^{(j)}(z)=\frac{-1}{\displaystyle  
z-\frac{\Omega_{{\rm R}j}^{2}}{z+\Omega_{{\rm R}j}^{2}~ 
m_{j}^{{\rm R}}(z)}}\,. 
\label{sechzehn} 
\end{equation} 
\end{mathletters}
Here the kernel $m_{j}^{{\rm R}}(z)$ is the $q\to 0$ limit of 
$m_{\rm R}(qj0,t)$. Carrying out the limit in the general formula for 
$m_{\rm R}(qj0,t)$ \cite{Franosch97c} and discretizing the wave-vector
integral afterwards, one finds: 
\begin{equation}
\label{siebzehn}
m_{j}^{\rm R}(t)={\cal F}\left[\Phi(kj\mu,t), \Phi_{k}(t)\right]
=\sum_{j~kl\nu}~ V_{kl\nu}^{j}~ \Phi(kl\nu,t)\Phi_{k}(t).
\end{equation}
The positive coupling coefficients $V_{kl\nu}^{j}$ 
are listed in Ref. \cite{Singh99}. 
After evaluation of $\Phi(kl\nu,t)$ and $\Phi_k(t)$ 
the correlators for the $M^{\prime}$
values of $k$, Eq. (\ref{siebzehn}) yields the
kernel $m^{\rm R}_{j}(t)$. Fourier-Laplace transformation gives $m^{\rm
R}_{j}(z)$ and Eq. (\ref{sechzehn}) provides
$C^{(j)}(z)$. Fourier-cosine transformation of the spectrum 
$C^{(j)\prime\prime}(\omega)$ leads to $C^{(j)}(t)$.

The theory shall be applied for a dumbbell consisting of two equal fused 
hard spheres of diameter $d$ and distance $\zeta d$ between the
centers. Thus, besides the packing fraction $\varphi$, there is the
elongation parameter $\zeta$ as the second control parameter specifying
the structure. The structure factor $S_{J}(q)$ and the corresponding pair
correlation functions $g_{J}(r)$ are evaluated within the Percus-Yevick
theory \cite{Franosch97b}. Figure \ref{zweiteAbb} exhibits the probability
distribution to find a solvent particle in the plane through the symmetry
axis of the dumbbell. The upper panel, calculated for $\zeta=0.80$, shows
a pronounced quadrupolar pattern extending over several shells. For the
small elongation $\zeta=0.33$, the lower panel shows that anisotropy is
almost lost from the third shell onwards. 
The calculations of the dynamics 
will be done for such moment of inertia that 
$v_{\rm R}/d=\sqrt{2}v/d$. The discretization will be done with
$M^{\prime}=50$ wave vectors with spacing $h^{\prime}=0.8$. 
The cutoff for the angular-momentum index is chosen 
as $l_{lo}=7$ for $\zeta=0.80$ and $\l_{lo}=5$ for $\zeta=0.33$.
The equation of motion (\ref{dreizehn}) is transformed to an
integro-differential equation in analogy to Eq. (\ref{einsa}) and then
solved by an algorithm similar to that used for the standard MCT
problem \cite{Singh99}. 

\section{Results}
\label{results}
\subsection{General features of reorientational relaxation}
Figure \ref{dritteAbb} demonstrates the transition scenario for the 
solute correlators for two 
representative wave vectors $q$ and three values of 
the angular momentum index $j$. The calculated correlators exhibit a 
very weak dependence on the helicity index $\mu$, and therefore only
the solutions for $\mu=0$ are shown. The wave vector $q=7.0$ is 
close to the structure-factor-peak position, and $q=10.6$ is near 
the first minimum of $S_q$. The correlator for $j=0$ is the 
probability distribution of the molecule's center-of-mass position, 
i.e. the analogue of the incoherent-intermediate scattering function 
for simple liquids: $\Phi(q00,t)=\Phi_q^s(t)$. Results for $j$=1 
and $j$=2 deal with the propagation of the dipole- and quadrupole-density 
fluctuations, respectively. The critical-decay 
curves, i.e. the solutions for 
$\varphi=\varphi_c$, organize the bifurcation pattern. 
They deal with the stretched decay towards the 
plateaus $f^c(qj0)$. If $\varphi$ increases above $\varphi_c$, the 
long-time limits $f(qj0)=\Phi(qj0,t\to\infty)$ increase above the 
plateau because the molecule gets more tightly localized in the frozen 
solvent. The $f(qj0)$-versus-$q$ curves are bell-shaped, since 
the molecules are localized with a nearly Gaussian probability 
distribution \cite{Franosch97c}. For $\varphi<\varphi_c$, the correlators 
exhibit a long-time decay from the plateau to zero, and this is 
the $\alpha$-process. The $\alpha$-decay time is the larger the 
smaller the wave vector $q$, while the 
$\alpha$-relaxation stretching increases with 
increasing $q$. The $q$-dependence of the 
relaxation features are similar as observed and explained previously 
for the tagged-particle correlator $\Phi_q^s(t)$ for simple liquids. 
Therefore, the following discussion shall be restricted to the 
($q=0$) limit, i.e. to the reorientational correlators $C^{(j)}(t)$.

Figure \ref{vierteAbb} exhibits representative decay curves $C^{(j)}(t)$ 
for the liquid state for two separation parameters, and 
Fig. \ref{fuenfteAbb} exhibits an extended set of susceptibility 
spectra 
$\chi^{(j)\prime\prime}(\omega)=\omega C^{(j)\prime\prime}(\omega)$. 
The plateaus and 
critical amplitudes shall be denoted by $f_j^c$ and $h_j$, respectively. 
They have been calculated from the mode-coupling functionals 
\cite{Franosch97c}, and some examples are listed in Tab. \ref{erstetabelle}. 
These parameters specify the leading-order asymptotic results for the 
$\beta$-relaxation process as explained in Sec. \ref{eqnsolv} for 
the solvent. The factorization theorem 
holds in analogy to Eq. (\ref{drei}) \cite{Franosch97c}:
\begin{equation}
\label{achtzehn}
C^{(j)}(t)=f_j^c + h_j~ G(t).
\end{equation}
The $\beta$-correlator $G$ is the same function as explained in 
connection with Eqs. (\ref{viera})--(\ref{sechs}) for the solvent. 
This implies for the critical correlator the asymptotic law
\begin{mathletters}
\label{neunzehn}
\begin{equation}
\label{neunzehna}
C^{(j)}(t)=f_j^c+h_j(t/t_0)^{-a}+{\cal{O}}((t/t_0)^{-2a})~ ;~ \sigma=0.
\end{equation}
The nonergodicity parameter of the glass state, $f_j=C^{(j)}(t\to\infty)$, 
exhibits the $\sqrt{\sigma}$-singularity
\begin{equation}
\label{neunzehnb}
f_j=f_j^c+h_j\sqrt{\sigma/(1-\lambda)}+{\cal{O}}(\sigma)~ ,~ \sigma\to0+~ ,
\end{equation}
and the $\alpha$-process initial decay is given by von Schweidler's law for
$t>t_\sigma$ and $\sigma\to0-$:
\begin{equation}
\label{neunzehnc}
C^{(j)}(t)=f_j^c~ [1-(t/\tilde{\tau}_\alpha^j)^b+{\cal{O}}_j
((t/\tilde{\tau}_\alpha^j))^{2b}]~ , \qquad 
\tilde{\tau}_\alpha^j=(f_j^c/h_j)^{1/b} ~ t_\sigma^{\prime} ~ .
\end{equation}
\end{mathletters}

Let us introduce two ad hoc time scales for the description of 
the liquid relaxation outside the transient regime. The center of the 
$\beta$-relaxation process, $\tau_\beta^j$, shall be defined as the 
time, where the correlator has decayed to the plateau: 
$C^{(j)}(\tau_\beta^j)=f_j^c$. The center of the $\alpha$-process 
shall be defined as the time, where the correlator has decayed to 50\% 
of the plateau: $C^{(j)}(\tau_\alpha^j)=f_j^c/2$. Some values 
are listed in Tab. \ref{zweitetabelle}, and open squares and 
circles mark these $\alpha$- and $\beta$-relaxation times, respectively,  
in Fig. \ref{vierteAbb}. The slowing down of the dynamics upon 
approaching the glass-transition point is reflected by the 
increase of the time scales with decreasing distance parameter 
$|\epsilon|$. The two step scenario emerges, because the ratio 
of the scales $\tau_\alpha^j/\tau_\beta^j$ increases as well. 
The $\alpha$-decay leads to the $\alpha$-peaks of the susceptibility 
spectrum, which are separated from the microscopic excitation peaks 
by a susceptibility minimum, as is demonstrated in Fig. \ref{fuenfteAbb}.
There, the $\alpha$-peak-maximum 
positions, $\omega_{\rm max}^{j}$, and the 
minimum positions, $\omega_{\rm min}^j$, decrease for 
$\varphi\to\varphi_c-$. The open squares and circles in 
Fig. \ref{fuenfteAbb} demonstrate, that 
$\omega_{\rm max}^j\approx 1/\tau_\alpha^j$ and 
$\omega_{\rm min}^j\approx 1/\tau_\beta^j$ as $|\epsilon|\to 0$. 
The two-step scenario implies that the ratio 
$\omega_{\rm max}^j/\omega_{\rm min}^j$ also decreases
upon approaching the glass-transition point. Thus, for 
$\varphi\to\varphi_c-$, the $\alpha$-peak gets more and more 
separated from the rest of the spectrum. In this limit, 
the plateau height is the relative area under the 
$\chi^{(j)\prime\prime}$-versus-$\log \omega$ curve \cite{Goetze91b}:
\begin{equation}
\label{zwanzig}
f_j^c=\int_{-\infty}^{\log \omega_{\rm min}^j}~ \chi^{(j)\prime\prime} 
(\omega)~ {\rm d} \log \omega / \int_{-\infty}^{\infty} 
\chi^{(j)\prime\prime} (\omega)~ {\rm d} \log \omega .
\end{equation}

Figure \ref{vierteAbb} demonstrates that for $t\lesssim 3$ 
the dynamics deals with oscillatory motion, i.e. with rotations and 
librations which are influenced by steric hindrance affects. If these 
effects would lead to some fast decay towards the correlator's 
long-time limit, one would find a white-noise 
low-frequency fluctuation spectrum: 
$C^{(j)\prime\prime} (\omega)\approx C^{(j)\prime\prime} (\omega=0)$.
Equivalently, one would 
obtain a regular low-frequency susceptibility spectrum varying  
linearly with $\omega$, 
$\chi^{(j)\prime\prime} (\omega) \propto \omega$, 
as is indicated schematically by the 
straight dashed-dotted line in the upper left 
panel of Fig. \ref{fuenfteAbb}. A linear susceptibility spectrum is 
obtained for the glass-spectra for $\omega\ll 1/t_\sigma$, since the 
correlators approach the limit $f_j$ exponentially for $t\gg t_\sigma$. 
This is shown by the ($\epsilon>0$) spectra in Fig. \ref{fuenfteAbb}. 
Such regular spectra are also found for the low-frequency wings of the 
$\alpha$-peaks, since the liquid correlators approach zero exponentially 
for $t\gg\tau_\alpha^j$. At the bifurcation point, however, the critical 
decay leads to a power law spectrum which, according to 
Eq. (\ref{neunzehna}), reads
\begin{equation}
\label{einundzwanzig}
\chi^{(j)\prime\prime}(\omega)=h_j~ \sin(a\pi/2)~ 
\Gamma(1-a)~ (\omega t_0)^a + {\cal{O}}((\omega t_0)^{(2a)}).
\end{equation}
For $t\ll t_\sigma$, the correlators follow the critical decay decay if
$|\sigma|$ is small. Therefore the spectra are approaching the 
asymptotic $\omega^a$-law for $1/t_\sigma\ll\omega\ll 1/t_0$, as is 
demonstrated for the ($x=4$) results in Fig. \ref{fuenfteAbb}. 
The stretching of the first relaxation step leads to the strong 
enhancement of the intensity of the spectral minimum 
$\chi_{\rm min}^j=\chi^{(j)\prime\prime}(\omega_{\rm min}^j)$ 
relative to any possible estimation of a white-noise-background 
spectrum. This enhancement also is 
exhibited by the experimental data reproduced 
in Fig. \ref{ersteAbb}.

Let us consider the probability density 
$P(\eta,t)=\langle \delta(\eta(t)-\eta) \rangle$ for the molecule's axis 
$\vec{e}(t)$ to have the projection $\eta(t)$ on its initial direction 
$\vec{e}$:
$\eta(t)=\vec{e}(t)\vec{e}$. Since 
$\delta(\eta(t)-\eta)=1/2+\sum_{j=1}^{\infty}~ (j+1/2)~ 
P_j(\eta) P_j(\eta(t))$, one gets
\begin{mathletters}
\label{zweiundzwanzig}
\begin{equation}
\label{zweiundzwanziga}
P(\eta,t)=1/2+\sum_{j=1}^{\infty}~ (j+1/2)~ 
P_j(\eta)~ C^{(j)}(t).
\end{equation}
Thus, knowledge of the set of $C^{(j)}(t)$, $j=1,2,...$, is equivalent 
to knowing $P(\eta,t)$. If the summation over $j$ is understood with the 
cutoff $l_{co}$, Eq. (\ref{zweiundzwanziga}) describes the evolution of the 
distribution with the initial value
\begin{equation}
\label{zweiundzwanzigb}
P(\eta,t=0)=1/2+\sum_{j=1}^{l_{co}}~ (j+1/2)~ P_{j}(\eta).
\end{equation}
\end{mathletters}
Figure \ref{sechsteAbb} exhibits results for the small distance 
parameter 
$-\epsilon=(\varphi_c-\varphi)/\varphi_c=0.001$ corresponding to $x=3$. 
The dotted lines exhibit $P(\eta,t=0)/10$, calculated with 
$l_{co}=7$ for $\zeta=0.80$, and $l_{co}=5$ for $\zeta=0.33$,
respectively. Within the dynamical window, where the 
leading-order result for the $\beta$-relaxation, Eq. (\ref{achtzehn}), 
applies, one gets
\begin{mathletters}
\label{dreiundzwanzig}
\begin{equation}
\label{dreiundzwanziga}
P(\eta,t)=P^c(\eta)+H(\eta)~G(t)~ ,
\end{equation}
\begin{equation}
\label{dreiundzwanzigb}
P^c(\eta)=1/2+\sum_{j=1}^{\infty}~ (j+1/2)~ f_j^c~ P_j(\eta) \, ; \qquad  
H(\eta)=\sum_{j=1}^{\infty}~ (j+1/2)~ h_j~ P_j(\eta)~ .
\end{equation}
\end{mathletters}
Thus the distribution relaxes towards the distribution 
$P^c(\eta)$, which is frozen for 
$\varphi=\varphi_c$. The relaxation does not exhibit any correlation
between changes in time described by $G(t)$, and variations with
angle described by $H(\eta)$. This is the scenario expected for 
relaxation due to dephasing in the random distribution of sizes 
and shapes of the cages producing steric hindrance for the 
rotations. For $\zeta=0.80$, the $\beta$ regime extends from 
$t=10$ to about $10^4$ as shown in Fig. \ref{vierteAbb}, and  the 
upper panel of Fig. \ref{sechsteAbb} exhibits the 
described phenomena for 
$t=10^2$ and $t=10^4$. The $\beta$-relaxation window is somewhat 
smaller for $\zeta=0.33$, as will be discussed in quantitative detail
below in connection with Fig. \ref{siebteAbb}. The dephasing 
relaxation for this case is demonstrated in Fig. \ref{sechsteAbb} 
for $t=10^2$ and $10^3$.

The beginning of the $\alpha$-relaxation process follows von 
Schweidler's law, Eq. (\ref{neunzehnc}). It is identical with the 
end of the $\beta$-process, and thus it is described within the
scenario based on Eqs. (\ref{dreiundzwanzig}). 
The most drastic difference between large- 
and small-elongation relaxation shows up for the $\alpha$-process 
outside the von-Schweidler regime.
For $\zeta=0.80$, the probability decreases monotonically if the
angle $\Theta$ of the axis increases from its initial value $\Theta=0$ 
to $\Theta=\pi$. This is shown in the upper panel of 
Fig. \ref{sechsteAbb} for $t\geq 5\cdot 10^5$. 
As time increases, 
the probability for $\eta\approx 1$ decreases, while it increases for 
$\eta\approx -1$. Thus, the relaxation towards the equilibrium  
distribution $P(\eta,t=\infty)=1/2$ is similar to what one would
expect for diffusion on a sphere. 
For $\zeta=0.33$, the correlators for odd $j$ decay faster than the 
corresponding correlators with the even index $(j+1)$. 
This is demonstrated in Fig. \ref{vierteAbb} and by the 
numbers $\tau_\alpha^j$ in Tab. \ref{zweitetabelle}. Therefore the 
$\alpha$-process consists of an intermediate time step leading 
to a probability distribution which is nearly symmetric with 
respect to the equator $\eta=0$. Only at later times, 
the symmetric distribution 
relaxes to the equilibrium one. Figure 
\ref{sechsteAbb} shows that, already for the rather short time $t=10^2$, 
$P(\eta,t)$ exhibits a minimum. For $t=10^3$, there is an overshooting 
effect of the probability for $\vec{e}(t)=-\vec{e}$: 
$P(\eta=-1, t=10^3)>0.5$;
and this effect increases if the time increases to $t=10^5$. Thus, 
the relaxation pattern is that expected for a random process of 
large-angle flips of the molecule's axis.

\subsection{Dipole-versus-quadrupole relaxation for 
strong steric hindrance}
The equations for the nonergodicity parameters \cite{Franosch97c} imply 
that the $f_j$ increase towards unity if the coupling coefficients in 
Eq. (\ref{vierzehn}) are increased towards infinity. 
For this strong-coupling limit, one 
derives from Eq. (\ref{zweiundzwanziga}) that 
$P(\eta,t)\to\delta(\eta-1)$. 
Because of continuity, for strong steric hindrance and 
for $t<t_\sigma$, $P(\eta,t)$ is a narrowly-peaked distribution 
centered around $\eta\approx 1$. Thus, one expects the expansion 
coefficients $f_j$ for not too large values of $j$ to form a smoothly 
decreasing sequence of $j$: $f_1>f_2>...$, $f_j\approx(f_{j-1}+f_{j+1})/2$. 
Table \ref{erstetabelle} demonstrates this result quantitatively for 
$\zeta=0.80$ and $\varphi=\varphi_c$:
\begin{mathletters}
\label{vierundzwanzig}
\begin{equation}
\label{vierundzwanziga}
f_1^c>f_2^c>f_3^c>f_4^c\, , \qquad \text{large $\zeta$}.
\end{equation}
In particular, the ratio $(f_1^c/f_2^c)$ of the relative strengths of 
the $\alpha$-peaks for the dipole relaxation, $f_1^c$, and for the 
quadrupole relaxation, $f_2^c$, is larger than unity. 
One cannot conclude quantitatively from $f_1^c/f_2^c$ 
the ratio 
$\chi_1^{\prime\prime}(\omega_{\rm max}^{1})/
\chi_2^{\prime\prime}(\omega_{\rm max}^{2})$ of the $\alpha$-peak heights, 
since the shapes of the spectra depend on $j$. 
However, Fig. \ref{fuenfteAbb} demonstrates that the two 
ratios are close to each 
other. One can also characterize the $\alpha$-peak height relative to 
the microscopic-peak height, 
$r_j=\chi^{(j)\prime\prime}(\omega_{\rm max}^{j})/
\chi^{(j)\prime\prime}(\omega_{\rm mic}^{j})$,
or relative to the minimum intensity, $r_j^{\prime}=
\chi^{(j)\prime\prime}(\omega_{\rm max}^j)/\chi^{(j)\prime\prime}
(\omega_{\rm min}^j)$. 
From Fig. \ref{fuenfteAbb}, one 
infers $r_1/r_2\approx 4$, and $r_1^{\prime}/r_2^{\prime}\approx 3$, 
i.e. the ($j$=1)-versus-($j$=2) enhancement 
effect appears even more pronounced.

According to Eq. (\ref{neunzehnb}), the nonergodicity parameters increase 
with increasing $(\varphi-\varphi_c)$. On the other hand, 
$1-f_j^c>f_j-f_j^c$. Therefore, $h_j$ must decrease if $f_j^c$ increases, 
so that the strongly coupled parameters $f_j$ leave the asymptotic 
regime for Eq. (\ref{neunzehnb}) for similar magnitudes of $\sigma$. 
Table \ref{erstetabelle} quantifies this result for $\zeta=0.80$.
In particular 
\begin{equation}
\label{vierundzwanzigb}
h_1<h_2\, , \qquad \text{large $\zeta$}.
\end{equation}
\end{mathletters}
The reasoning assumes $f_j$ to be large, and thus it cannot be 
applied for too large $j$. There is some $j_0$, so that $h_j$ 
decreases with increasing $j$ for $j>j_0$. Within the frequency window, 
where the leading-order asymptotic law for the critical 
decay is valid, Eq. (\ref{einundzwanzig}), one derives 
an enhancement of the ($j$=2) spectrum 
relative to the ($j$=1) spectrum, since:
\begin{equation}
\label{fuenfundzwanzig}
\chi^{(2)\prime\prime}(\omega)/\chi^{(1)\prime\prime}(\omega)
=h_2/h_1. \qquad 1/t_\sigma\ll\omega\ll 1/t_0~ .
\end{equation} 
The dotted lines $c_1$ and $c_2$ in the upper panels of 
Fig. \ref{fuenfteAbb} demonstrate this result. 

For a strongly near-($\eta$=1)-peaked probability 
distribution $P(\eta,t)$, one can 
approximately replace averages of functions of $\eta$ by the 
functions of the average $\langle \eta \rangle$. Thus, Lebon 
et al. concluded: $f_j=P_j(f_1)$ \cite{Lebon97}. Specializing to 
$\varphi=\varphi_c$, one quantifies the sequence of 
$f_j^c$ in terms of its first value $f_1^c$:
\begin{mathletters}
\label{sechsundzwnzig}
\begin{equation}
\label{sechsundzwnziga}
f_j^c=P_j(f_1^c)~ , \qquad \zeta\to\infty~ .
\end{equation} 
Substituting into Eq. (\ref{neunzehnb}) and specializing to $\sigma\to 0+$,
one can also quantify the sequence of $h_j$ by the 
first term $h_1$: 
\begin{equation}
\label{sechsundzwanzigb}
h_j=P_j^{\prime}(f_1^c)~h_1~ , \qquad \zeta\to\infty~ .
\end{equation} 
\end{mathletters} 
From Tab. \ref{erstetabelle} one infers, that for $\zeta=0.80$ the 
error of Eq. (\ref{sechsundzwnziga}) for $j=2~ (3,4)$ is as small as 
$0.1\%~ (3\%,7\%)$, and Eq. (\ref{sechsundzwanzigb}) is obeyed for 
$j=2~ (3,4)$ within $5\%~ (22\%, 45\%)$. 

The strong nonlinear couplings of the structural-relaxation modes require 
that all correlators enter the first relaxation step, the second relaxation 
step, and the equilibrium state nearly at the same respective time. 
This is demonstrated in Fig. \ref{dritteAbb} for $\zeta=0.80$. The most
striking manifestation of the coupling effect occurs at the center of
the $\beta$-relaxation window for $\sigma\to 0-$. In this case, the 
factorization theorem, Eq. (\ref{achtzehn}), is valid. All
correlators cross their plateau at the same time, say $\tau_\beta$, where 
$\tau_\beta$ is the zero of the $\beta$ correlator $G(t)$. Because of the 
scaling law, Eq. (5), one gets the result $\tau_\beta=\hat{t}_{-} t_\sigma$, 
i.e. 
\begin{equation}
\label{siebenundzwanzig}
\tau_\beta^{j}=\hat{t}_{-}t_\sigma~ , \qquad \sigma\to 0_{-}~ .
\end{equation}
Here, $t_\sigma$ is the scale from Eq. (\ref{sechsa}), 
and $\hat{t}_{-}$ is the zero of the 
master function: $g_{-}(\hat{t}_{-})=0$. 
For the HSS it reads $\hat{t}_{-}=0.704$
\cite{Franosch97}. The open and full circles in Fig. \ref{vierteAbb} 
show that the asymptotic Eq. (\ref{siebenundzwanzig}) 
is obeyed very well for 
$\zeta=0.80$. Since the $\alpha$-processes 
of $C^{(1)}(t)$ and $C^{(2)}(t)$ start at the same time 
$\hat{t}_{-}t_\sigma$ and reach zero nearly at the 
same time, one expects from 
$C^{(1)}(\hat{t}_{-}t_\sigma)=f_1^c>f_2^c=C^{(2)}(\hat{t}_{-}t_\sigma)$ 
that the decay time for $C^{(1)}$ is larger than that for
$C^{(2)}$:
\begin{equation}
\label{achtundzwanzig}
\tau_\alpha^{1}>\tau_\alpha^{2}\, , \qquad \text{large $\zeta$}.
\end{equation}
Furthermore, the $C^{(1)}(t)/f_1^c$-versus-$\log t$ plot is somewhat 
steeper than the corresponding graph for $j$=2. This means that the 
stretching is larger for the ($j$=2) $\alpha$ process than 
for the ($j$=1) $\alpha$ process. If one interpolates the 
decay functions by a Kohlrausch law, $C^{(j)}(t)/f_j^c\approx 
\exp\left[ -(t/\tau_\alpha)^{\beta_j} \right]$, the stretching exponent for
$j$=1 is larger than that for $j$=2:
\begin{equation}
\label{neunundzwanzig}
\beta_1>\beta_2\, , \qquad \text{large $\zeta$}.
\end{equation}
Stretching can also be quantified by the width $w$ at half height 
of the $\alpha$-peak of the susceptibility spectrum. For 
$\zeta=0.80$ our model yields for $j=1(2,3,4)$ $w=1.16(1.25,1.37,1.50)$ 
decades. The Kohlrausch processes leading to the same $w$ require 
stretching exponents $\beta=0.99(0.90,0.82,0.74)$.

The derivation of the inequality for the time scales can be put on
a quantitative level by combining Eq. (\ref{neunzehnc}) 
with the 
two inequalities in Eqs. (\ref{vierundzwanzig}). One gets 
in analogy to Eq. (\ref{achtundzwanzig}): 
$\tilde{\tau}_\alpha^1>\tilde{\tau}_\alpha^{2}$. The $\alpha$-relaxation
law for the $C^{(j)}(t)$ holds in analogy to Eq. (\ref{sieben}):
$C^{(j)}(t)=\tilde{C}_j(\tilde{t})$. 
If the shape function 
$\tilde{C}_j(\tilde{t})/f_j^c$ would be independent of $j$, the ratio 
$\tau_\alpha^{1}/\tau_\alpha^{2}$ would be equal to the ratio 
$\tilde{\tau}_\alpha^1/\tilde{\tau}_\alpha^2
=[f_1^c h_2/h_1 f_2^c]^{1/b}$. But the latter 
is about $2.1$ times larger than 
$\tau_\alpha^{1}/\tau_\alpha^{2}$. 

\subsection{Dipole-versus-quadrupole relaxation for weak steric 
hindrance}
There are two universal phenomena which are relevant for a
discussion of the dynamics for weak steric hindrance. The first one
concerns the limit $\zeta=0$ of the center-of-mass correlator
$\Phi(j=0~ \mu=0~ q, t)=\Phi_q^s(t)$, which is identical to the 
tagged-particle-density correlator of the larger of the two 
spheres forming the dumbbell. If the radius of this sphere, say $d_1$, 
is of the same order or larger than the radius $d$ of the solvent 
spheres, the steric hindrance is very effective. In this case, 
$\Phi_q^s(t)$ exhibits the canonical bifurcation 
scenario if $\varphi$ crosses $\varphi_c$, as was discussed 
comprehensively in Ref. \cite{Fuchs98}. This implies 
that for $d_1\gtrsim d$ the ($j=0$) correlators exhibit only a smooth  
$\zeta$-dependence for $\zeta$ decreasing to zero.
A side remark shall be added to this conclusion. If the ratio 
of the diameters $d_1/d$ of a sphere moving in a glass of hard spheres 
decreases towards zero, there occurs a percolation
transition at some critical value $(d_1^c/d)$. This is a type-A 
transition, i.e. a bifurcation where the 
Lamb-M{\"o}ssbauer factor decreases continuously to zero
for $(d_1-d_1^c)$ approaching zero from above 
\cite{Franosch94,Goetze91b}.
Because of continuity, it is obvious that for a 
dumbbell built of sufficiently small spheres, 
$d_1<d_1^c$, there will be a type-A transition if the elongation 
$\zeta$ decreases to some critical value $\zeta^{*}>0$. If 
$\zeta$ crosses $\zeta^{*}$, the dynamics changes from one 
dealing with molecules localized in the hard-sphere glass 
to one dealing with delocalized molecular motion. This small-$\zeta$ 
phenomenon for small molecules is not considered in this paper.

The second universal phenomenon deals with a type-A transition 
resulting from the fact, that for top-down 
symmetrical molecules the MCT equations of motion of the 
even-$j$ correlators decouple from the odd-$j$ ones 
\cite{Franosch97c,Singh99}. 
The even-$j$ 
correlators couple to the function $\Phi_q^s(t)$, and thus the 
conventional transition scenario of this correlator enforces 
the same for all other correlators with even $j$.
However, such coupling does not exist for odd $j$. 
For large $\zeta$, this results in no considerable effect. But 
all coupling coefficients in the equations of motion 
approach zero for odd $j$ if $\zeta$ tends to zero.
Consequently, for all $\varphi>\varphi_c$ there is some critical 
elongation $\zeta_c(\varphi)$ for a type-A transition. 
For the studied model $\zeta_c(\varphi)<\zeta_c(\varphi_c)=0.296$ 
\cite{Franosch97c,Franosch98b}. 
Choosing $\zeta$ sufficiently close to
$\zeta_c(\varphi)$, it can happen that for odd $j$: $f_j<f_{j+1}$ or
even $f_j<f_{j+3}$ \cite{Franosch97c}. 
The transition at $\zeta_c(\varphi)$ 
shall not be studied in this paper. 
For the demonstration of the small-($\zeta-\zeta_c(\varphi_c)$) 
phenomena, the value $\zeta=0.33$ has been
chosen so large that the canonical sequence 
for the plateau values in the ($q=0$) limit, Eq. (\ref{vierundzwanziga}), 
is not violated, as is quantified in Tab. \ref{erstetabelle}. 
But it is chosen so small, that the precursor effects
of the type-A transition seriously 
influence the results for the dumbbell dynamics. 
Thereby, the results are also representative for such cases, where the 
type-A transition singularity is avoided \cite{Franosch94} due to a 
weak breaking of the top-down symmetry of the solvent-solute 
interaction.

The even-$j$ correlators show the conventional behavior. Therefore, the 
discussion of their trends with decreasing $\zeta$ for fixed $j$ can
be held analogously to that given in Sec. IIIB 
for the trends with increasing $j$ for fixed large 
$\zeta$. Thus one understands that the ($j$=2) $\alpha$-process for
$\zeta=0.33$ is weaker, faster, and more stretched than that for 
$\zeta=0.80$, as it is demonstrated in Figs. \ref{vierteAbb} and 
\ref{fuenfteAbb}. 
The halfwidth of the $\alpha$-peak for $j=2(4)$ is $w=1.66(1.86)$ decades 
as for a Kohlrausch process with exponent $\beta=0.67(0.59)$. 
Notice in particular from Fig. \ref{vierteAbb}, that the 
$\beta$-relaxation scale $\tau_\beta^{2}$ for the $x=3$-result is
close to the $\zeta$- and 
$j$-independent number $\hat{t}_{-}t_\sigma$ from
Eq. (\ref{siebenundzwanzig}). For $x=2$, the asymptotic formula is 
obeyed reasonably, but the preasymptotic corrections are larger for
$\zeta=0.33$ than for $\zeta=0.80$.

The most obvious precursor of the type-A transition is 
the suppression of the plateau values $f_j$ for odd $j$. This leads 
to a violation of the rule $(f_1+f_3)/2\approx f_2$, as is quantified in 
Tab. \ref{erstetabelle}. The general qualitative reasoning
from Sec. IIIB explains, that the suppression of $f_1$ is connected with
an enhancement of $h_1$: 
$h_1(\zeta=0.33)/h_1(\zeta=0.80)\approx 15$. The amplitude 
$h_1$ is given by the resolvent of the so-called stability matrix, and 
at a type-A transition the resolvent exhibits a pole 
\cite{Franosch94,Goetze91b}. Hence 
$h_1(\zeta\to\zeta_c(\varphi))/h_2(\zeta\to\zeta_c(\varphi))\to\infty$,  
and the regular trend, Eq. (\ref{vierundzwanzigb}), is reversed:
\begin{equation}
\label{einunddreissig}
h_1>h_2\, , \qquad \text{small $\zeta$}.
\end{equation}
For our example one infers from
Tab. \ref{erstetabelle} that $h_1/h_2\approx 4.2$. According to 
Eq. (\ref{fuenfundzwanzig}), the critical spectrum for the dipole 
relaxation is considerably larger than that for the quadrupole 
relaxation, as is demonstrated by the dotted lines in the lower 
two panels of Fig. \ref{fuenfteAbb}. 

Combining Eq. (\ref{einunddreissig}) with von Schweidler's law,
Eq. (\ref{neunzehnc}), one concludes that the 
$C^{(1)}(t)$-versus-$\log t$ curve crosses its plateau $f_1^c$ 
much steeper than the 
$C^{(2)}(t)$-versus-$\log t$ curve. This is illustrated in the
lower panel of Fig. \ref{vierteAbb}. Hence the $\alpha$-relaxation 
of the ($j$=1) response is faster than the one of the ($j$=2) response:
\begin{equation}
\label{zweiunddreissig}
\tau_\alpha^{1} < \tau_\alpha^{2}\, , \qquad \text{small $\zeta$}.
\end{equation}
Again the order for large $\zeta$, Eq. (\ref{achtundzwanzig}), 
is reversed. From Tab. \ref{zweitetabelle}, one infers for $x=3$: 
$\tau_\alpha^{1}/\tau_\alpha^{2}=0.12$.
Accordingly, the $\alpha$-peak positions for the ($x=3$) spectra for
$j$=1 and $j$=2 in the lower panels of Fig. \ref{fuenfteAbb} 
differ by about one 
order of magnitude. For the ratio of the von Schweidler scales in
Eq. (\ref{neunzehnc}), one gets 
$\tilde{\tau}_\alpha^1/\tilde{\tau}_\alpha^2=[h_2f_1^c/h_1f_2^c]^{1/b}\to 0$
for $\zeta\to\zeta_c$, and this identifies the 
smallness of the ratio $\tau^1/\tau^2$ 
as a precursor of the type-A transition.
The preceding discussion is valid more generally and explains that 
all the odd-$j$ correlators decay faster than the 
nearby even-$j$ ones. As a result, the 
probability distribution $P(\eta,t)$ approaches first an even 
function of $\eta$, before the equilibrium value $0.5$ is obtained, 
as is demonstrated in the lower panel of Fig. \ref{sechsteAbb}.

\subsection{$\beta$-relaxation scaling}

The factorization theorem for the $\beta$-relaxation, 
Eq. (\ref{achtzehn}), means that the rescaled correlators
$c^{(j)}(t)=(C^{(j)}(t)-f_j^c)/h_j$ are given  
independently from $j$ by the $\beta$-correlator 
$G(t)$ of the solvent. The latter obeys the scaling law, specified by 
Eqs. (\ref{viera})-(\ref{sechsa}). 
For fixed rescaled time $\hat{t}=t/t_\sigma$, the cited formulas 
deal with the results correctly up to order
$\sqrt{|\sigma|}$
\cite{Goetze91b}.
The leading corrections are of order $|\sigma|$, and they explain
the range of validity of the leading results for separations $\epsilon$ 
\cite{Franosch97}.
Figure \ref{siebteAbb} demonstrates these statements. 
On a $10\%$-accuracy 
level the leading-order results describe $4\% (18\%; 45\%; 20\%)$ 
of the decay of the correlators
around the plateau for $\zeta=0.80$, $j=1 (\zeta=0.80, j=2;
\zeta=0.33, j=1; \zeta=0.33,j=2)$.
These decay intervals are indicated in Fig. \ref{vierteAbb} 
by vertical lines. For $\epsilon=-0.001$, the corresponding dynamical
window extends from about $t=10$ to about $10^5$, 
while it extends from about $t=3$ to about $t=100$ for 
$\epsilon=-0.01$. This discussion requires a reservation: 
The corrections to the scaling
results can lead to such a violation of Eq. (\ref{siebenundzwanzig}),
which appears as an offset of the plateau \cite{Franosch97}. This
offset can be noted in the lower panel of Fig. \ref{siebteAbb} for the
odd-$j$ results. 
The good description of the $\beta$-decay of the ($\zeta=0.33$) 
results for odd $j$ holds only after a correction of 
the offset. 
For $t\gtrsim 10^4$, the correction effects cause the
$c^{(j)}(t)$ for $\zeta=0.80$ to differ from $G(t)$; 
one infers from Fig. \ref{siebteAbb} that the $c^{(j)}(t)$ increase with
increasing $j$. The general results for the theory of the
corrections imply, that then also the $c^{(j)}(t)$ increase with $j$ for
$t\lesssim 10$ \cite{Franosch97}. The $c^{(j)}(t)$-versus-$\log
t$ curves do not intersect for $\hat{t}_{-}t_\sigma$ but they touch
each other as is demonstrated in the upper panel of 
Fig. \ref{siebteAbb}. Corresponding results also hold for
$\zeta=0.33$ after the mentioned offset is eliminated.

Equations (\ref{fuenf}) and (\ref{achtzehn}) lead to the scaling
law for the susceptibility spectra:
$\chi^{(j)\prime\prime}(\omega)/h_j=c_{\sigma} \hat{\chi}_{\pm}
(\omega t_{\sigma})$. The master spectra
$\hat{\chi}_{\pm}(\hat{\omega})=\hat{\omega}
g_{\pm}^{\prime\prime}(\hat{\omega})$ are given by the Fourier-cosine
transform $g_{\pm}^{\prime\prime}(\hat{\omega})$ of the master functions
$g_{\pm}(\hat{t})$. The master spectrum for the glass state describes the
crossover from a regular spectrum for small rescaled
frequencies, $\hat{\chi}_{+}(\hat{\omega}\ll 1)\propto
\hat{\omega}$, to the critical spectrum at large rescaled
frequencies, $\hat{\chi}_{+}(\hat{\omega}\gg 1)\propto
\hat{\omega}^{a}$. It deals with the knee exhibited by the spectra
for $\epsilon>0$ and $x=3,4$ in Fig. \ref{fuenfteAbb}. The master
spectrum for the liquid describes the crossover from the
von-Schweidler high-frequency tail of the $\alpha$-peak, 
$\hat{\chi}_{-}(\hat{\omega}\ll 1)\propto 1/\hat{\omega}^{b}$, to the
critical decay for large rescaled frequencies, 
$\hat{\chi}_{-}(\hat{\omega}\gg 1)\propto \hat{\omega}^{a}$. 
The results describe in the small-$\sigma$ limit the 
$\beta$-relaxation minimum as it can be seen in Fig. \ref{fuenfteAbb} 
for the $x=3$ and $x=4$ results. In particular, 
the factorization theorem explains why the spectral minima 
$\omega_{\rm min}$ are
located at the same position independently of $j$ and $\zeta$. The
leading-order formulas imply
$\omega_{\rm min}=\hat{\omega}_{\rm min}/t_{\sigma}$, 
where $\hat{\omega}_{\rm min}$
denotes the minimum of the master spectrum $\hat{\chi}_{-}$. For
the hard-sphere system, one gets $\hat{\omega}_{\rm min}=1.56$
\cite{Franosch97}.

Obviously, the $\beta$-relaxation scaling laws can describe the
susceptibility minimum only for such small distance parameters,
for which $\omega_{\rm min}$ is located in that frequency window where the
($\sigma=0$) spectrum exhibits the asymptotic $\omega^{a}$-law,
Eq. (\ref{einundzwanzig}). Figure \ref{fuenfteAbb} shows that for
the model under study this window is restricted to
$\omega<0.01$. This means, that $\omega_{\rm min}$ has to be located
about three decades below the peak of the microscopic susceptibility 
spectrum. For $\omega>0.01$, the critical spectrum is modified by
crossover effects to the transient dynamics. The
susceptibility minimum with $\omega_{\rm min}>0.01$ is due to the
crossover of the $\alpha$-peak tail to the microscopic excitation 
spectrum;
it cannot be discussed by the universal asymptotic laws for the MCT
bifurcation. 
One concludes from Fig. \ref{fuenfteAbb}, that 
$|\epsilon|<10^{-2}$ needs to be satisfied in order to apply 
the $\beta$-scaling laws for the model under study.

\subsection{$\alpha$-relaxation scaling}
The $\alpha$-relaxation scaling law 
reads for the reorientational correlators in analogy to 
Eq. (\ref{sieben}):
\begin{equation}
\label{neuneu}
C^{(j)}(t)=\tilde{C}_j(\tilde{t})~ , 
\qquad \tilde{t}=t/t_{\sigma}^{\prime}~ .
\end{equation}
The $\epsilon$-independent master function $\tilde{C}_j$ 
obeys as initial condition the von Schweidler law:
$\tilde{C}^{(j)}(\tilde{t})=f_j^c-h_j\tilde{t}^b+{\cal{O}}(\tilde{t}^{2b})$.
The superposition principle for the susceptibility spectra reads 
$\chi^{(j)\prime\prime}(\omega)=
\tilde{\chi}^{(j)\prime\prime}(\tilde{\omega})$ with 
$\tilde{\omega}=\omega t_\sigma^{\prime}$ denoting the rescaled frequency. 
The $\epsilon$-independent master spectrum is given by the 
Fourier-cosine transform of the master correlators 
$\tilde{\chi}^{(j)\prime\prime}(\tilde{\omega})=
\tilde{\omega} \tilde{C}^{(j)\prime\prime}(\tilde{\omega})$. 
Consequently, the above defined 
$\alpha$-relaxation time scales $\tau_\alpha^{j}$ 
and susceptibility-maximum positions $\omega_{\rm max}^{j}$ read 
\begin{equation}
\label{dreiunddreissig}
\tau_\alpha^j=\tilde{t}^j t_\sigma^{\prime}~ , \qquad 
\omega_{\rm max}^{j}=\tilde{\omega}^j/t_\sigma^{\prime}~ ,
\end{equation}
where $\tilde{t}^j$ is defined by $\tilde{C}^{(j)}(\tilde{t}^j)=f_j^c/2$ 
and $\tilde{\omega}^j$ denotes the peak frequency of 
$\tilde{\chi}^{(j)\prime\prime}(\tilde{\omega})$. The scaling law implies 
that a representation of $C^{(j)}(t)$ as a 
function of the rescaled time $t/\tau_\alpha^j$ should 
superimpose correlators for different distance parameters $\epsilon$ on the
common curve $\tilde{C}^{(j)}(\tilde{t}/\tilde{t}^j)$. Asymptotic 
validity means that the $\log (t/\tau_\alpha^j)$ interval, where the
scaling law is obeyed, expands to arbitrary size for $\epsilon\to 0$. A 
corresponding statement holds for the representation of the susceptibility
peaks as functions of the rescaled frequency. The corrections to the 
leading-order asymptotic laws are 
the larger, the larger the critical amplitude 
$h_j$ is \cite{Franosch97,Fuchs98}. Figures \ref{fuenfteAbb} and 
\ref{achteAbb} demonstrate, that the described scenario for the 
evolution of the $\alpha$-process is valid for $\zeta=0.80$, and also 
for $\zeta=0.33$ in the case $j$=2. 
For strong steric hindrance, the $\alpha$-scaling law works for 
larger values of $(\varphi_c-\varphi)$, than the $\beta$-scaling law.
This is so, because the
leading corrections to the $\alpha$-scaling law are of relative 
size ${\cal{O}}(|\epsilon|)$, while they are of relative size 
${\cal{O}}(\sqrt{|\epsilon|})$ for the $\beta$-scaling law \cite{Franosch97}.

Figure \ref{achteAbb} demonstrates a drastic ($j$=1)-versus-($j$=2) 
effect of the $\alpha$-scaling for $\zeta=0.33$. The dipole 
correlators do not exhibit the superposition principle for 
$|\epsilon| > 10^{-4}$, nor do the correlators for the other odd values 
of $j$. For $j=1$ the plateau emerges only for extremely small values of 
the distance parameter 
$|\epsilon|\leq 10^{-4}$. The $\alpha$-peak heights of the dipole 
spectra decrease with decreasing $|\epsilon|$ in Fig. \ref{fuenfteAbb} 
in contradiction to the scaling-law prediction. This anomaly 
is caused by the large size of the critical 
amplitude $h_1$, which was explained in connection with 
Eq. (\ref{einunddreissig}).
More precisely, it is caused by the large percentage of the decay of 
$C^{(1)}(t)$ described by the $\beta$-scaling law as 
is indicated by the vertical lines in Fig. \ref{vierteAbb}. 
To formulate this observation quantitatively, let us remember that the 
decay of the correlator near the plateau is described by 
Eqs. (\ref{fuenf}), (\ref{sechs}) and (\ref{achtzehn}):
$C^{(j)}(t)=f_j^c+ h_j \sqrt{|\sigma|} g_{-}(t/t_\sigma)$. 
The master function $g_{-}(\hat{t})$ for small positive 
values and all negative ones is 
well approximated by:
$\hat{g}(\hat{t})=-B\hat{t}^b+B_1/(B\hat{t}^b)$. Here, $B_1$ is 
determined by the exponent parameter $\lambda$ and for our 
solvent model reads $B_1=0.431$ \cite{Franosch97}. Thus one gets for 
$C^{(j)}(t)\lesssim f_j^c$ within the window for the validity of the 
$\beta$-relaxation scaling law:
\begin{mathletters}
\label{vierunddreissig}
\begin{equation}
\label{vierunddreissiga}
C^{(j)}(t)=f_j^c-\sqrt{|\sigma|} h_j 
\{ B(t/t_\sigma)^b-B_1/[B(t/t_\sigma)^b] \}~ .
\end{equation}
The leading corrections to this formula can explain the possible 
offset of $f_j^c$ or, equivalently, of the scales $t_\sigma$ 
\cite{Franosch97}, which was noticed above in connection with 
Fig. \ref{siebteAbb} for $\zeta=0.33$. 
Equation (\ref{vierunddreissiga}) can be
rewritten as 
$C^{(j)}(t)=[f_j^c-h_j \tilde{t}^b] + h_j |\sigma| B_1/\tilde{t}^b$.
Here, the bracket is the $\alpha$-scaling-law description of the 
initial part of the $\alpha$-process, and the term proportional 
to $B_1$ is the leading correction. 
The correction term to the $\alpha$-scaling law deals with that part 
of the $\beta$-process below the plateau, which is not given by 
the von Schweidler's large-$\hat{t}$ asymptote. 
Therefore, one can write the for the $\alpha$-process for 
not too large values of rescaled time $\tilde{t}$:
\begin{equation}
\label{vierunddreissigc}
C^{(j)}(t)=\tilde{C}^{(j)}(\tilde{t})+h_j |\sigma| B_1/\tilde{t}^b~ .
\end{equation}
\end{mathletters}
The analogue of this formula was shown in Ref. \cite{Franosch97} 
to describe the evolution of the $\alpha$-process of the 
density correlators of the HSS perfectly for 
$|\epsilon|\leq 0.1$. It was also shown, that the corresponding spectrum 
describes the susceptibility peak to increase above the scaling-law 
constant $\tilde{\chi}^{\prime\prime}(\tilde{\omega}_{\rm max})$ if the 
separation $|\epsilon|$ increases from $10^{-2}$ to $10^{-1}$. 

In Fig. \ref{neunteAbb}, the evolution of the ($j$=1) $\alpha$-process 
for small steric hindrance is reexamined. Instead of rescaling the time
with the ad-hoc scale $\tau_\alpha^1$, the 
theoretically motivated scale $t_\sigma^{\prime}$ is chosen. One 
recognizes, that the found scenario does not exhibit any 
qualitative peculiarity anymore, compared to what is presented in 
Fig. \ref{achteAbb} for $\zeta=0.33$ and $j$=2. 
The ($j$=1)-versus-($j$=2) anomaly is identified as an anomaly of the 
size of the corrections only. 
In the case of the small elongation, the distance parameter 
$|\epsilon|$ has to be taken almost two orders of magnitude smaller 
for $j$=1 in order to render the corrections to the $\alpha$ scaling 
as small as found for $j$=2. For $|\epsilon|\gtrsim 10^{-3}$, even 
including the leading corrections to von Schweidler's law, 
one can explain the relaxation from 
the plateau only up to some offset in the time scale. This is 
demonstrated in Fig. \ref{neunteAbb} by the dotted lines for $x=2,3$.

Two remarks concerning tests of the second scaling law shall be added. 
The definition of the time scale $\tau_\alpha^j$ used in 
Fig. \ref{achteAbb} was arbitrary.
Let us consider more general definitions to be parameterized by a number
$k>1$ and denoted as $\tau_k$. The subscripts $\alpha$ and $j$ shall be
dropped for the sake of simplicity, and the definition shall be:
$C^{(j)}(\tau_k)=f_j^c/k$. If the scaling law is valid, one finds in 
analogy to Eq. (\ref{dreiunddreissig}): 
$\tau_k=\tilde{t}_k t_\sigma^{\prime}$.
Here, $\tilde{t}_k$ is 
defined by $\tilde{C}^{(j)}(\tilde{t}_k)=f_j^c/k$. In this case, the 
choice of $k$ is irrelevant, since the ratio of two different scales is 
$\epsilon$-independent: 
$\tau_{k_1}/\tau_{k_2}=\tilde{t}_{k_1}/\tilde{t}_{k_2}$.
However, if preasymptotic corrections are present, the scales are not 
equivalent. The range of validity of the superposition 
principle expands from large to small rescaled times. This 
follows from Eq. (\ref{vierunddreissigc}) and is demonstrated in 
Fig. \ref{neunteAbb}. One gets for $k_1<k_2$:
\begin{equation}
\label{fuenfunddreissig}
\tau_{k_1}/\tilde{t}_{k_1}<\tau_{k_2}/\tilde{t}_{k_2}<t_\sigma^{\prime}
\end{equation}
For a detection of the superposition principle for an as large as
possible value of $|\epsilon|$, one should therefore choose an as large as 
possible value of $k$ for the rescaling procedure. Thereby, the 
artificial crossing point of the rescaled curves at $t=\tau_k$ is
suppressed as much as possible. Otherwise, one introduces a time 
scale $\tau_k$ for the characterization of a decay process 
which cannot be characterized by a single scale. The outcome of this  
ill-defined procedure is demonstrated in the lower left panel 
in Fig. \ref{achteAbb}. In this case, $\tau_\alpha^1=\tau_{k=2}$ is a 
parameter extracted from the correlator which, according
to Fig. \ref{siebteAbb}, is adequately to be specified by the 
two scales $c_\sigma$ and $t_\sigma$ of the $\beta$-relaxation 
scaling law. The dashed line in the 
inset of Fig. \ref{neunteAbb} demonstrates explicitly, 
that the scale $\tau_\alpha^1$ does not exhibit the asymptotic 
behavior for $|\epsilon| \geq 10^{-3}$. The asymptotic law  
$\tau_\alpha^1=\tilde{t}_2 t_\sigma^{\prime}$ is followed only for 
$\epsilon\leq 10^{-4}$. 

The second remark concerns the determination of the exponent $\gamma$ 
entering the power-law behavior for the $\alpha$-relaxation time 
scale, as specified by Eqs. (\ref{sechsb}) and (\ref{dreiunddreissig}).
These results are based on the validity of 
the scaling law \cite{Goetze91b}.
Therefore, one cannot appeal to MCT if one fits power laws for scaling 
times for cases where the scaling law is violated.
The dashed line in the 
inset in Fig. \ref{neunteAbb} demonstrates, that the scale
$\tau_\alpha^1$ for $|\epsilon|\geq 10^{-3}$ can be fitted well by 
a power law for a two-decade variation of the distance parameter 
$|\epsilon|$. The identified effective exponent 
$\gamma^{\prime}<\gamma$ describes the variation of $\tau_{\alpha}^{1}$ 
over three orders of magnitude; but nevertheless $\gamma^{\prime}$ 
has no well defined meaning for the discussion of our model.

\section{Conclusions} 
\label{conclusions} 
Solving the MCT equations of motion for the dynamics of a hard-sphere 
dumbbell moving in a hard-sphere liquid, first-principle 
results have been obtained 
for the evolution of the glassy dynamics of the reorientational 
degrees of freedom of a molecule. It was found that one has to distinguish 
between two scenarios, namely between strong steric hindrance as found 
for large elongations $\zeta$ of the dumbbell, and weak steric hindrance 
as found for small elongations.

For strong steric hindrance, the mode-coupling coefficients 
for the reorientational degrees of freedom 
in Eq. (\ref{vierzehn}) are of the same order as the ones entering 
Eq. (\ref{zwei}) for the description of the translational degrees 
of freedom of the solvent. The dependence of the various 
parameters on the angular-momentum index $j$ is similar to the 
dependence on the wave vector $q$. 
One has to view $j(j+1)$ as the analogue of $q^2$.
While the $q$-dependence reflects the decomposition
of the direct solute-solvent correlations in plane waves 
the $j$-dependence reflects the decomposition in spherical harmonics.
Hence one finds, that 
all reported results on the $j$-dependence of the reorientational 
correlators $C^{(j)}(t)$ are similar --- and can be explained in a 
similar manner --- as known from the previous work on the 
tagged-particle-density correlators $\Phi_q^s(t)$ in simple liquids 
\cite{Fuchs98,Fuchs92b}.  
In particular, it was 
shown that with increasing $j$ the $\alpha$-peak-strength 
parameters $f_j^c$, Eq. (\ref{vierundzwanziga}), the 
$\alpha$-relaxation-time scales $\tau_\alpha^{j}$, 
Eq. (\ref{achtundzwanzig}), and the stretching exponents 
$\beta_j$, Eq. (\ref{neunundzwanzig}), decrease. 
These findings reproduce qualitatively the three general 
differences between dielectric-loss and 
depolarized-light-scattering spectra, which were discussed in 
Sec. I in connection with Fig. \ref{ersteAbb}.
Because of Eq. (\ref{einsa}), the relaxation of the correlator 
$\Phi$ follows that of the kernel $m$. Therefore, the $\alpha$-relaxation 
time scale of the ($q\to 0$) density fluctuations, say $\tau^0_{\alpha}$, 
is larger than the corresponding scale of the longitudinal 
elastic modulus $m_{q=0}(z)$, 
say $\tau^m_{\alpha}$. For strong steric hindrance, the decay of the cage 
is the prerequisite for the reorientation of the molecule, and therefore 
$\tau^0_{\alpha} < \tau^2_{\alpha}$.  Thus one expects the fourth general 
feature of the $\alpha$ relaxation listed in the introduction: 
$\tau^2_{\alpha}/\tau^m_{\alpha}>1$. For our model one gets for 
$\epsilon=-0.01$: $\tau^m_{\alpha}=130$, $\tau^0_{\alpha}=240$, 
$\tau^2_{\alpha}=920$. The ratio $\tau^2_{\alpha}/\tau^m_{\alpha}\approx 7$ 
is of the same order as cited in Sec. I for PC and Salol.

The mode-coupling coefficients in Eq. (\ref{vierzehn}) decrease to zero 
if $j$ tends to infinity. Thus the solutions for large $j$ are sums of
many small terms, which are not strongly correlated. Each term exhibits 
the short-time von Schweidler-law behavior for the $\alpha$-relaxation:
$C^{(j)}(t)-f_j^c\propto (t/t_\sigma^{\prime})^b$.
Therefore, one expects for $C^{(j)}(t)$ 
the characteristic function of the stable L{\'e}vy 
distribution, $\exp\left[(-t\Gamma_j)^b\right]$ \cite{Goetze92}. 
For the density correlators of the solvent, Fuchs has 
worked out the limit behavior for the $\alpha$-relaxation 
master function for $q \to \infty$ 
and showed how the Kohlrausch law with $\beta=b$ arises \cite{Fuchs94}. 
We suspect that a similar derivation can be done for the 
reorientational correlators. Therefore, we conjecture that the 
sequence of Kohlrausch exponents 
$\beta_1>\beta_2>\beta_3 ...$ converges towards the von Schweidler 
exponent $b$. 
Molecular-dynamics-simulation data for a model of water have 
been interpreted consistently within the standard MCT scenario 
\cite{Gallo96,Sciortino96,Sciortino97}. In particular, the 
$C^{(j)}(t)$ exhibit the conventional behavior \cite{Fabbian98}. 
One concludes that water exhibits 
strong-steric-hindrance effects. Therefore it is reinsuring that the
sequence of the first five Kohlrausch exponents $\beta_j$ decreases 
with increasing $j$ monotonously towards the von Schweidler exponent 
\cite{Sciortino99}.
A further general result, namely the increase in the initial part of the
series of critical amplitudes, Eq. (\ref{vierundzwanzigb}), is also 
found in the simulation data for $j\leq 3$ 
\cite{Sciortino99}.
  
Figure \ref{ersteAbb} exhibits as full lines the 
($j$=1) and ($j$=2) spectra calculated for 
$\zeta=0.80$. The lines for $x=1$ and $x=2$ 
are the ones discussed in Fig. \ref{fuenfteAbb}, and the other 
two refer to $x=1.33$ and $x=1.67$, respectively. In order to 
transfer the MCT results, which are calculated with ad hoc units 
specified in Sec. II, to the units used by the experimentalists, 
one has to introduce three scales. The first and second scale transfer
the calculated dimensionless normalized spectra 
$\chi^{(j)\prime\prime}(\omega)$ for $j$=1 and $j$=2 to the units used
by the experimentalists for their 
dielectric-loss and depolarized-light-scattering spectra, 
respectively. The third scale shifts our frequency scale to the 
GHz scale. In the double-logarithmic
representation, the first two scales define an overall vertical shift 
of the diagrams in Fig. \ref{fuenfteAbb}, while the third scale defines 
a horizontal shift of the figures. Intending to compare data for 
PC for different temperatures with the MCT results for different 
packing fractions $\varphi$, one gets a 
mapping of the $T$-scale onto the $\varphi$-scale via Eq. (\ref{elf}).
The result is shown as an inset in Fig. \ref{ersteAbb}. The inset also 
includes the point with coordinates of the critical packing fraction of
the solvent $\varphi_c$ and the critical temperature $T_c=180K$. This 
value for $T_c$ was determined for PC by analyzing neutron scattering data 
\cite{Boerjesson90}, and has recently been corroborated 
in an MCT analysis of various other PC experiments \cite{Goetze00}.
Our results in Fig. \ref{ersteAbb} describe the evolution of the two 
types of PC spectra semi-quantitatively. In particular, the extrapolation 
of the $T$-$\varphi$-relation yields a reasonable estimation of the 
critical temperature for that system, which is demonstrated through the 
dashed line in the inset. 
There is no obvious reason 
why the studied dilute solution of hard symmetric dumbbells in a 
hard-sphere solvent should produce spectra, which are similar to the 
data for PC. We consider the found similarities 
to a large extend as accidental. The theoretical curves are added in 
Fig. \ref{ersteAbb} with the mere intention to justify the conclusion:
the model studied in this paper 
and our choice of parameters are relevant for achieving 
an understanding of experiments in glass-forming molecular liquids.

In order to further corroborate the preceding conclusion, let us consider 
Fig. \ref{zehnteAbb}. The data points exhibit a susceptibility 
spectrum of PC measured by incoherent-neutron-scattering 
spectroscopy for the wave vector $q=1.3 {\rm \AA}^{-1}$ 
\cite{Wuttke99}. A remarkable feature of the $\alpha$-peak spectrum 
compared to the spectra shown in Fig. \ref{ersteAbb} is that it is less 
pronounced relative to the spectrum of the microscopic excitation band 
and that is is more stretched. The two dashed lines in Fig. \ref{zehnteAbb} 
exhibit the spectra for the center-of-mass correlator 
$\Phi_{q}^{s}(t)=\Phi(q,j=0,\mu=0,t)$ for $q=7.4$ in order to emphasize, 
that this leading approximation for the scattering function 
cannot easily explain the experimental findings. 
The shown $\alpha$-peaks of $\Phi_q^s$ have a half width of $w=1.34$ decades 
as produced by a Kohlrausch process with exponent $\beta=0.84$.
The scattering function $F_q(t)$
is a sum over the contributions of the molecule's constituents and hence 
it is a superposition of the density correlators for all angular 
momentum indices $j$. For the symmetric dumbbell one gets up to some 
normalization constant \cite{Franosch97c}
\begin{equation}
\label{neutronengleichung}
F_{q}(t)=\sum_j~ (2j+1)~ b_{j}(q~\zeta/2)^{2}~ \Phi(q,j,0,t)~ ,
\end{equation}
where $b_{j}(z)$ are the spherical Bessel functions. The full lines in
Fig. 10 exhibit the spectra for $F_q(t)$ for $q=7.4$. 
The $\alpha$-peaks have a halfwidth of $w=1.61$ decades as produced by 
a Kohlrausch law with stretching exponent $\beta=0.69$. 
The frequency was rescaled as
explained in connection with Fig. \ref{ersteAbb} and the scale for the 
spectra was adjusted to meet the one of the data. Comparison of the 
full line with the dashed one for $x=2$ shows the features 
distinguishing the $\alpha$-processes of $F_{q}^{\prime\prime}(\omega)$ 
from that of $\Phi_{q}^{s \prime\prime}(\omega)$. The result calculated 
for $x=1$ shows that the finding for our model semiquantitatively 
accounts for the $\alpha$-peak data.

Some side remarks considering the 
comparisons in Figs. \ref{ersteAbb} and \ref{zehnteAbb}  
might be useful. 
A schematic-model analysis of the PC data 
gave the exponent parameter $\lambda\approx 0.75$ \cite{Goetze00}, 
in good agreement with the values found from analyses of the susceptibility 
minima with the $\beta$-relaxation scaling laws 
\cite{Schneider99,Du94,Wuttke99}.
The value is close to the result $\lambda\approx 0.74$ for the hard-sphere 
system, Eq. (\ref{neun}).
This accident ensures that the master function for the susceptibility 
minimum and the values of all anomalous exponents of PC agree 
within the experimental uncertainties with 
the corresponding quantities of the model studied in this paper; and 
this is a prerequisite of a successful fit.
Accidently, the ratio of the $\alpha$-relaxation times 
$\tau_\alpha^1/\tau_\alpha^2$ noted in Tab. \ref{zweitetabelle} for
$\zeta=0.8$ is only a bit larger than the ratio of the 
$\alpha$-peak-maximum positions of PC,
$\omega_{\rm max}^2/\omega_{\rm max}^1$, exhibited in Fig. 1; and this is 
another request for a reasonable fit.
Since the ratio decreases with decreasing $\zeta$, some $\zeta<0.80$ 
could be chosen to reproduce the specified 
($j$=1)-versus-($j$=2) effect quantitatively.
Nevertheless, it is remarkable that the fit in Fig. \ref{ersteAbb} 
reproduces the ratios of 
$\alpha$-peak maximum intensity to $\beta$-minimum 
intensity $\chi^{(j)\prime\prime}(\omega_{\rm max})/
\chi^{(j)\prime\prime}(\omega_{\rm min})$
reasonably well for both values of $j$. Neither it is trivial, that the 
model reproduces reasonably 
the ($j$=1)-versus-($j$=2) effect for the stretching. 

A new liquid-glass-transition scenario is predicted which is 
referred to as the regime of weak steric hindrance for reorientational 
motion. It is characterized by ($j$=1)-versus-($j$=2) effects, 
more generally by odd-$j$-versus-even-$j$ effects, which are 
qualitatively different from the results described above as 
strong-steric-hindrance results. The new scenario occurs, if 
precursor phenomena of a type-A-transition between two 
nonergodic states strongly influence the asymptotic results
for the conventional MCT bifurcation. The scenario appears if the 
particle interactions deviate not too strongly from spherical symmetry, 
e.g., if a linear molecule exhibits only small deviations from a 
top-down symmetry and if there are not too large elongations. 
Six features characterize the weak-steric-hindrance scenario.
First (i), the plateaus $f_j^c$ for the reorientational correlators
for odd $j$ are suppressed in comparison 
to what one would expect by interpolating 
or extrapolating the values for nearby even-$j$ plateaus 
(Tab. \ref{erstetabelle}).
Most importantly (ii), the critical amplitude $h_1$ is larger than 
$h_2$, Eq. (\ref{einunddreissig}), so that the canonical ordering of 
the $h_j$ for small $j$, Eq. (\ref{vierundzwanzigb}), is reversed. 
Third (iii), the percentage of the decay of the reorientational 
correlators $C^{(j)}(t)$, which can be explained by the 
leading- plus next-to-leading-order asymptotic formulas for the 
$\beta$-relaxation is larger for $j$=1 than for $j$=2; as is indicated 
by the vertical lines in Fig. 4. The structural relaxation 
of the reorientations is dominated by large-angle flips (iv), as 
shown in Fig. \ref{sechsteAbb} for the dumbbell with elongation 
$\zeta=0.33$. The $\alpha$-relaxation-time scale for $j$=1 is 
smaller than for $j$=2, Eq. (\ref{zweiunddreissig}), (v) so that the 
canonical order of the $\alpha$-relaxation scales, 
Eq. (\ref{achtundzwanzig}), is reversed. This can cause the 
$C^{(j)}(t)$-versus-$\log t$ graphs for $j$=1 to cross the graphs  
for $j$=2, as is shown for the $\zeta=0.33$ results in Figs. \ref{dritteAbb} 
and \ref{vierteAbb}. Finally (vi), for distance parameters 
$|\epsilon|\geq 10^{-3}$, where the conventional 
$C^{(j)}(t)$-versus-$\log t/(\tau_\alpha^j)$ plot exhibits the 
$\alpha$-relaxation scaling law for $j$=2, the correlators for $j$=1 do 
not show the validity of the superposition principle, as is demonstrated 
in Fig. \ref{achteAbb}. Nor does the scale
$\tau_\alpha^1$, defined as the time for a 50\% decay of the 
$\alpha$-relaxation correlator, exhibit the power law behavior 
with the correct exponent $\gamma$ as is shown in the inset of 
Fig. \ref{neunteAbb}. 

A side remark concerning a molecular-dynamics study of the 
evolution of glassy dynamics in a Lennard-Jones-dumbbell liquid by
K{\"a}mmerer et al. \cite{Kaemmerer97,Kaemmerer98,Kaemmerer98b} 
might be in order. It was reported that the correlators 
dealing with translational degrees of freedom and also for the ones 
for the reorientational dynamics for angular index $j\neq 1$ could 
be interpreted qualitatively within the universal asymptotic MCT 
formulas. However, the evolution of the dipole correlators did not 
fit into the standard MCT pattern. It was found that $h_1>h_2$ and 
$\tau_\alpha^1<\tau_\alpha^2$. A drastic violation of the 
$\alpha$-scaling law was noted quite similar to what is 
exhibited in the lower left panel of Fig. \ref{achteAbb}. The 
scale $\tau_\alpha^1$ exhibited a deviation from 
the asymptotic law $\epsilon^{-\gamma}$, but a fit by 
$\tau_{\alpha}\propto \epsilon^{-\gamma^{\prime}}$ with 
$\gamma^{\prime}$ as discussed in the inset of 
Fig. \ref{neunteAbb} was possible. 
These simulation results for $j$=1 
differ from those for water simulations 
\cite{Sciortino96,Sciortino97,Fabbian98,Sciortino99}
as well as from the experimental findings for 
propylene carbonate quoted in Fig. \ref{ersteAbb}.
However, they 
agree with the features (ii), (v), and (vi) specified in the preceding 
paragraph. Moreover, the property (iv) concerning the large angle flips 
is also obtained in Ref. \cite{Kaemmerer97}. Accidently, the $x=2$ results 
in the lower panel of Fig. \ref{fuenfteAbb} show that the minimum 
position of the $j$=1 spectrum exceeds that of the $j$=2 spectrum by 
nearly one order of magnitude, in agreement with the corresponding finding 
in Ref. \cite{Kaemmerer98b}. 
Furthermore, the $\alpha$-peak variation with $x$ shown in the lower 
left panel of Fig. \ref{fuenfteAbb} is in qualitative agreement with that 
reported in Ref. \cite{Kaemmerer97}.
In view of these observations it 
does not seem impossible, that the scenario studied in Refs. 
\cite{Kaemmerer97,Kaemmerer98,Kaemmerer98b} fits into the 
framework of the ideal MCT. However, 
it is not clear, whether or not the results 
of Ref. \cite{Kaemmerer97,Kaemmerer98,Kaemmerer98b} can be explained 
by our theory for type-A precursors of a dilute solution of 
molecules. First, the simulation results for the 
dumbbell liquid do not exhibit a particular decrease of $f_1^c$ 
relative to $f_2^c$. 
Second, the 
$\beta$-relaxation scaling has not been documented for the dumbbell 
liquid and so it is unclear, whether or not the feature (iii) holds 
for that case.

Summarizing, it shall be emphasized that all qualitative features 
for the evolution of the structural relaxation studied in this paper 
have been explained by means of the formulas for the leading-asymptotic 
expansions and their leading-order-correction formulas for the bifurcation 
scenario. In this sense, these asymptotic 
formulas can be considered as the essence 
of MCT. However, in order to explain the characteristic 
($j$=1)-versus-($j$=2) differences for the relaxation patterns,  
it is necessary to also understand the general trends of the 
nonuniversal parameters with variations of wave-vector $q$ and
angular-momentum index $j$. And this requires to use 
MCT as a microscopic theory based on the knowledge of the equilibrium 
structure.

\bigskip 
\begin{acknowledgements} 
We thank the authors of Refs. \cite{Schneider99,Du94,Wuttke99} for 
the permission to use their data files and M. Fuchs for  
many stimulating discussions. 
We kindly thank F. Sciortino for the permission to cite his unpublished 
simulation results for the reorientational dynamics of water. 
We acknowledge gratefully helpful critique and suggestions for 
improvements of the manuscript written to us by H.Z. Cummins, R. Pick, and 
R. Schilling. This work was supported by Verbundprojekt BMBF 03-G05TUM.  
\end{acknowledgements}


\begin{table}
\caption{Plateau values $f_j^c$ and critical amplitudes $h_j$}
\begin{tabular}{ccccc}
 &\multicolumn{2}{c}{$\zeta=0.80$}&\multicolumn{2}{c}{$\zeta=0.33$}\\
 \tableline
j   & $f_j^c$ & $h_j$ & $f_j^c$ & $h_j$ \\ 
1   & 0.943 & 0.13 & 0.303 & 1.94 \\ 
2   & 0.835 & 0.35 & 0.286 & 0.46 \\ 
3   & 0.701 & 0.55 & 0.052 & 0.46 \\ 
4   & 0.540 & 0.68 & 0.006 & 0.13 \\  
\end{tabular}
\label{erstetabelle}
 \end{table}
 
\begin{table}
\caption{Time scales $\tau_\alpha^j$ and $\tau_\beta^j$}
\begin{tabular}{ccccc}
 &\multicolumn{2}{c}{$\zeta=0.80$}&\multicolumn{2}{c}{$\zeta=0.33$}\\
 \tableline
  & $x=2$ & $x=3$ & $x=2$ & $x=3$ \\
$\tau_\alpha^1$ & $2.75\times 10^3$ & $7.65\times 10^5$ & 
$9.29\times 10^1$ & $6.51\times 10^3$ \\
$\tau_\alpha^2$ & $9.21\times 10^2$ & $2.56\times 10^5$ & 
$1.85\times 10^2$ & $5.39\times 10^4$ \\
$\tau_\alpha^3$ & $4.40\times 10^2$ & $1.20\times 10^5$ & 
$5.98\times 10^1$ & $4.43\times 10^3$ \\
$\tau_\alpha^4$ & $2.41\times 10^2$ & $6.43\times 10^4$ & 
$6.55\times 10^1$ & $1.74\times 10^4$ \\
\tableline 
$\tau_\beta^1$  & $1.37\times 10^1$ & $5.42\times 10^2$ & 
$3.33\times 10^1$ & $8.11\times 10^2$ \\
$\tau_\beta^2$  & $1.31\times 10^1$ & $5.34\times 10^2$ & 
$1.03\times 10^1$ & $4.80\times 10^2$ \\
$\tau_\beta^3$  & $1.32\times 10^1$ & $5.32\times 10^2$ & 
$2.53\times 10^1$ & $7.24\times 10^2$ \\
$\tau_\beta^4$  & $1.25\times 10^1$ & $5.22\times 10^2$ & 
$1.00\times 10^1$ & $4.74\times 10^2$
\label{zweitetabelle}
\end{tabular}
\end{table}
 
\begin{figure}[tbp] 
\bigskip 
\caption[fig1]{
Susceptibility spectra $\chi^{\prime\prime}$ of propylene carbonate 
(PC, symbols) and solutions obtained for a symmetric hard-sphere 
dumbbell with elongation $\zeta=0.80$ immersed in a hard-sphere solvent 
(full lines, see text for details). The symbols represent dielectric-loss 
spectra measured by Schneider et al. \cite{Schneider99} 
(upper panel) and depolarized-light-scattering spectra 
of Du et al. \cite{Du94} (lower panel) for temperatures as 
indicated. The full lines are calculated for the distance 
parameter $\epsilon=(\varphi-\varphi_c)/\varphi_c=-10^{-x}$ 
with $x=1, 1.33, 1.67$ and $2$ for angular momentum index 
$j$=1 and $2$, respectively. 
Computed frequencies have been rescaled by a factor of 10 to meet 
the experimental GHz scale. The calculated susceptibilities 
have been divided by 2.8 for the $j$=1 case in order to normalize 
the spectra at $\omega/2\pi=2$GHz.
The inset exhibits packing fraction 
$\varphi$ versus temperature $T$ 
for which the spectra are fitted.
Here, the critical value of the hard-sphere system, 
$\varphi_c=0.516$, 
corresponding to the critical temperature of PC, 
$T_c\approx 180$K, was added. The dashed line 
demonstrates the extrapolation from the found $\varphi$-$T$-mapping 
to $T_c$.}
\label{ersteAbb} 
\end{figure} 
\bigskip 
 
\begin{figure}[tbp] 
\caption[fig2]{
Angular dependent solute-solvent pair-distribution function 
$g(\vec{r},\vec{\Omega})$,  
calculated within the Percus-Yevick theory, for a top-down 
symmetric solute molecule consisting of two equal fused hard 
spheres with elongation $\protect\zeta=0.8$ (upper panel) and  
$\protect\zeta=0.33$ (lower panel). The shown $x$--$z$ plane 
contains the molecule axis. Grey corresponds to 
$g(\vec{r},\vec{\Omega})\approx 1$, dark and white 
areas show regions with higher and lower probability to 
find a solvent particle, respectively. 
The cut through the dumbbell is shown hatched. 
The diameter $d$ of each sphere is chosen 
to match that of the surrounding solvent particles. 
The unit of length is chosen here and in all 
following figures such that $d=1$. The packing 
fraction of the hard-sphere solvent is at the critical value 
$\varphi_c=0.516$. 
$g(\vec{r},\vec{\Omega})$ was approximated using a 
Legendre-polynomial expansion with angular momentum indices up to
$j=16$.}
\label{zweiteAbb} 
\end{figure} 
\bigskip 
 
\begin{figure}[tbp] 
\caption[fig3]{
Correlators $\Phi$ for the wave vectors $q=7.0$ and $10.6$, 
elongations $\zeta=0.80$ (upper two panels) 
and $\zeta=0.33$ (lower two panels), angular indices $j=0,1,2$, and 
helicity index $\mu=0$ as functions of the logarithm of the time 
$t$. The unit of time is chosen here and in all following figures 
such that the thermal velocity $v$ of the solvent is unity. 
Correlators are shown as full lines for $j=0,2$ and as 
dashed lines for $j$=1. 
The solutions at the critical packing fraction are marked 
by a {\it c} and are shown in dotted. The packing fractions are 
parameterized as $(\varphi-\varphi_c)/\varphi_c=\epsilon=\pm 10^{-x}$,  
and $x=1,2,3,4$ was chosen. Solutions for the glass states, 
$\epsilon>0$, are only shown for $\zeta=0.80$, $j$=1.
Correlators are truncated where necessary to avoid overcrowding of the
figure.}
\label{dritteAbb} 
\end{figure} 
\bigskip 
 
\begin{figure}[tbp] 
\caption[fig4]{ 
Reorientational correlators $C^{j}(t)$ for $j$=1 and $j$=2 for the two 
elongations $\zeta=0.80$ and $\zeta=0.33$ as function of $\log_{10} t$. 
The solutions at the critical point are shown in dotted and are marked by 
$c_j$. The plateau values $f_j^c$ are marked by horizontal lines. The 
distance parameter is chosen as 
$\epsilon=(\varphi-\varphi_c)/\varphi_c=-10^{-x}$ with $x=3$ (slower 
decay) and $x=2$ (faster decay). Open circles and open squares mark the 
characteristic time scales $\tau_\beta^j$ and $\tau_\alpha^j$ for the 
$\alpha$- and $\beta$-process, respectively. The full circles 
and squares mark the time scales $0.704~ t_\sigma$, with $t_{\sigma}$ from 
Eq. (\ref{sechsa}), and $t_\sigma^{\prime}$ from Eq. (\ref{sechsb}), 
respectively.
The vertical lines indicate the decay interval described by the asymptotic 
formulas for the $\beta$-process (see text, cf. Fig. \ref{siebteAbb}). }
\label{vierteAbb} 
\end{figure} 
\bigskip 
 
\begin{figure}[tbp] 
\caption[fig5]{
Double-logarithmic presentation of the susceptibility spectra 
$\chi^{(j)\prime\prime}(\omega)=\omega C^{(j)\prime\prime}(\omega)$ 
for angular-momentum indices $j$=1 and $j$=2 for elongations 
$\zeta=0.80$ and $\zeta=0.33$. Spectra for the critical packing fraction 
$\varphi=\varphi_c$ are shown in dotted and are marked by $c_j$. 
The distance parameters are $\epsilon=\pm 10^{-x}$ with $x$ as given 
in the panels. In the upper left panel, a regular 
susceptibility spectrum, $\chi^{\prime\prime}\propto \omega$, 
corresponding to a 
white-noise fluctuation spectrum, is indicated by a dashed-dotted 
straight line of slope unity. The open circles and squares mark the 
frequencies $1/\tau_\beta^j$ and $1/\tau_\alpha^j$,  
characterizing the $\beta$- and $\alpha$-relaxation process, respectively.
The full circles and full squares mark the frequencies 
$1/t_\sigma$ and $1/t_\sigma^{\prime}$, respectively.}
\label{fuenfteAbb} 
\end{figure} 
\bigskip 

\begin{figure}[tbp] 
\caption[fig6]{
Evolution of the probability density $P(\eta,t)$ to find at time 
$t$ the molecular axis $\vec{e}(t)$ with projection $\eta(t)$ 
onto its initial direction. 
The dotted lines are the initial distributions, 
Eq. (\ref{zweiundzwanzigb}), downscaled by a factor of $10$.
The oscillations around $P(\eta,t)=0$ are due to restricting the 
infinite sum over angular-momentum indices in Eqs. (\ref{zweiundzwanzig})
to $j\leq7$ (upper panel) and $j\leq5$ (lower panel), respectively.}
\label{sechsteAbb} 
\end{figure} 
\bigskip 

\begin{figure}[tbp] 
\caption[fig7]{
The full lines exhibit the reorientational correlators rescaled to 
$c^{(j)}(t)=[C^{(j)}(t)-f_{j}^{c}]/h_{j}$ for two distance parameters 
$\epsilon$ and the angular-momentum indices $j=1-4$. The dashed 
lines show the $\beta$-correlator $G(t)=c_\sigma g_{-}(t/t_\sigma)$ 
of the hard-sphere system, obtained from 
Eqs. (\ref{viera}), (\ref{fuenf}), and (\ref{sechsa}). }
\label{siebteAbb} 
\end{figure} 
\bigskip 
 
\begin{figure}[tbp] 
\caption[fig8]{
Reorientational correlators $C^{j}(t)$ for $j=1,2$ and 
$\zeta=0.80$ and $\zeta=0.33$ for various distance parameters 
$\epsilon=-10^{-x}$, presented as 
functions of $\log_{10}(t/\tau_\alpha^j)$. The 
$\alpha$-relaxation-time scale $\tau^j_{\alpha}$ is defined by 
$C^{(j)}(\tau_\alpha^j)=f_j^c/2$. The horizontal lines 
indicate the plateaus $f_j^c$.}
\label{achteAbb} 
\end{figure} 
\bigskip 
 
\begin{figure}[tbp] 
\caption[fig9]{
Dipole correlator $C^{(1)}(t)$ of the dumbbell with small elongation 
$\zeta=0.33$ and distance parameters $\epsilon=-10^{-x}$ for
$x=2$--$5$ as functions of the logarithm of the rescaled time 
$\tilde{t}=t/t_\sigma^{\prime}$ (light full lines). 
Here $t_\sigma^{\prime}$ is the 
second critical time scale, Eq. (\ref{sechsb}). 
The heavy full line is the $\alpha$-relaxation 
master function $\tilde{C}^{(1)}(\tilde{t})$. The dotted lines 
show the leading-order $\alpha$-scaling result plus the 
leading correction term according to Eq. (\ref{vierunddreissigc}). 
The inset exhibits in a double-logarithmic plot 
$t_\sigma^{\prime}$ (full squares) and the ad-hoc scaling time 
$\tau_\alpha^{1}$ (open squares) for $x=1$--$5$. The full 
straight line with slope $\gamma=2.46$ exhibits the power-law 
formula for the hard-sphere system, Eq. (\ref{sechsb}). 
The dashed line interpolates the open squares for 
$x=1,2,3$ with an effective power law exponent $\gamma^{\prime}=1.65$,
while the dotted line is the asymptotic small-$\epsilon$ result 
for $\tau_\alpha^{1}$.}
\label{neunteAbb} 
\end{figure} 
\bigskip 

\begin{figure}[tbp] 
\caption[fig10]{
Susceptibility spectrum $\chi^{\prime\prime}$ of PC 
as measured by incoherent neutron scattering 
\cite{Wuttke99} for $q=1.3{\rm \AA}^{-1}$ and $T=285$K 
(circles). The solid lines exhibit the 
neutron-scattering response of the discussed MCT model, 
and the dashed lines are the mere center-of-mass 
contributions for packing fractions corresponding to 
$x=1$ and $2$. The computational wave vector is $q=7.4$. As done in 
Fig. \ref{ersteAbb}, a rescaling of the theoretical 
frequencies by a factor of 10 was chosen to match the scale of the 
experiment. The normalized theoretical spectra have 
been rescaled by a factor of $1.1$.
}
\label{zehnteAbb} 
\end{figure} 
\bigskip 
\end{document}